\pdfoutput=1
\documentclass[preprint,12pt]{elsarticle}
\usepackage[T1]{fontenc}
\usepackage[utf8]{inputenc}
\usepackage[a4paper,margin=2.2cm]{geometry}
\usepackage{amsmath,amssymb,graphicx,booktabs}
\usepackage[table]{xcolor}
\definecolor{famirtint}{HTML}{E8F3F9}
\definecolor{famirtablerule}{HTML}{79868E}
\newcommand{\famirtablestyle}{%
  \sffamily\fontsize{8}{10.5}\selectfont
  \renewcommand{\arraystretch}{1.16}%
  \setlength{\tabcolsep}{2pt}%
  \setlength{\heavyrulewidth}{0.7pt}%
  \setlength{\lightrulewidth}{0.4pt}%
  \arrayrulecolor{famirtablerule}}
\usepackage{microtype}
\usepackage{placeins}
\usepackage{float}
\usepackage[hidelinks]{hyperref}
\usepackage{url}
\journal{Computer Methods in Applied Mechanics and Engineering}
\biboptions{sort&compress}
\begin{document}
\begin{frontmatter}
\title{FAMIR: Future-aligned latent dynamics for complex-flow forecasting}
\author[tju]{Yuxuan Chen}
\author[tju]{Yan Wang}
\author[tju]{Haotian Cheng}
\author[tju]{Hongna Zhang\corref{cor1}}
\ead{hongna@tju.edu.cn}
\author[yzu]{Wenhua Zhang}
\author[heu]{Yuke Li}
\author[tju]{Xiaobin Li}
\author[tju]{Fengchen Li}
\cortext[cor1]{Corresponding author.}
\affiliation[tju]{organization={School of Mechanical Engineering, State Key Laboratory of Engines, Tianjin University},
  city={Tianjin}, postcode={300350}, country={China}}
\affiliation[yzu]{organization={School of Petroleum Engineering, Yangtze University},
  city={Wuhan}, postcode={430100}, country={China}}
\affiliation[heu]{organization={College of Shipbuilding Engineering, Harbin Engineering University},
  city={Harbin}, postcode={150001}, country={China}}
\begin{abstract}
Learned flow surrogates are applied recursively, so errors in phase, amplitude and transport accumulate as each prediction feeds the next. During training, however, the realized next state is known. We propose Future-Aligned Motion Inference for Recursive Physical Forecasting (FAMIR), which uses this realized future to shape the signal that drives the update, not only the predicted state. On the latent grid of a frozen autoencoder, a future-conditioned teacher and a history-only student infer motion tensors that condition one shared residual transition. Posterior--Prior Motion Alignment (PPMA) jointly aligns the teacher (posterior) and student (prior) motions, prediction-fed multistep objectives train the student on its own rollouts, and only the student is deployed. On viscoelastic channel flow, Rayleigh--B\'enard convection and planetary shallow-water flow, FAMIR achieves lower complete-prefix normalized root-mean-square error than three direct latent predictors of comparable size and a delay-embedded linear reference at ten steps (1.003\%, 1.055\%, 0.620\%) and twenty steps (1.661\%, 1.683\%, 1.036\%). Its advantage is largest at the first step in the viscoelastic flow and grows with lead time in the other flows. Over forty-step forecasts, FAMIR keeps fluctuation profiles and probability densities closer to the reference than the direct predictors. Without the teacher, the motion-conditioned student already outperforms the direct predictors; in the shallow-water flow, prediction-fed rollouts are decisive and joint alignment is far more effective than distillation toward a fixed teacher (1.036\% against 1.874\%). Fixed-state interventions show that the trained update relies on the state-specific motion, and the deployed student matches the one-step accuracy of the future-informed teacher.

\end{abstract}
\end{frontmatter}
\section{Introduction}

Accurate prediction of complex flow evolution underpins scientific analysis, design exploration, flow control and any task that requires many repeated simulations. High-fidelity solvers remain the reference for these tasks, but their cost limits parameter sweeps, forecast ensembles and real-time use. Machine learning has become an established route to fast flow surrogates \cite{brunton2020machine,vinuesa2022enhancing,kochkov2021machine,thuerey2020deep}. A learned surrogate is useful only if it reproduces more than plausible snapshots: it must retain local structures, coherent interfaces and large-scale transport while the flow passes through changing dynamical states.

Surrogates of this kind are usually deployed recursively. Each prediction becomes part of the input history for the next step, so small errors in phase, amplitude or transport are carried forward and compound over the forecast. The model is also queried on histories that it has itself generated, which differ from the observed histories used in standard one-step training. How strongly these effects act depends on the flow. Polymer conformation in viscoelastic channel flow (VF) couples velocity to material memory and forms sharp stress gradients; Rayleigh--B\'enard convection (RBC) organizes temperature into intermittent plumes and thin boundary layers; planetary shallow-water flow (PlanetSWE) combines localized vortices with planetary-scale transport on a sphere. The common question is how to represent such fields compactly and how to learn a state update that remains accurate when its inputs are its own predictions.

Reduced-order modeling addresses the representation. Proper orthogonal decomposition (POD) constructs energy-ranked linear coordinates, and dynamic mode decomposition (DMD) fits a linear evolution operator to snapshot sequences \cite{berkooz1993proper,schmid2010dynamic,rowley2017model,taira2017modal}. Koopman-operator theory places such operators on a common footing \cite{mezic2013analysis}, and delay coordinates enrich the information available to them \cite{arbabi2017ergodic,brunton2017chaos}. Nonlinear alternatives learn the coordinates from data: autoencoders capture structures beyond a linear modal basis, and learned embeddings can be paired with sparse or approximately linear dynamics \cite{murata2020nonlinear,eivazi2020deep,lusch2018deep,otto2019linearly,brunton2016discovering,fukami2021sindy}. Neural predictors have also been used to advance learned latent states in time, from latent-space fluid simulation to multiscale effective dynamics \cite{wiewel2019latent,vlachas2022multiscale}. Recurrent networks have forecast a low-order model of turbulent shear flow \cite{srinivasan2019predictions}, and a three-dimensional convolutional autoencoder combined with a long short-term memory network has forecast minimal turbulent channel flow in its latent space \cite{nakamura2021cnnlstm}. Variational and operator-learning models further separate compression from temporal modeling \cite{solerarico2024beta,kontolati2024learning}. These works motivate a spatially structured latent state, but a good representation does not by itself determine how that state should be advanced.

Neural operators and attention models increase the expressiveness of the update map. Operator-learning methods such as the Fourier neural operator (FNO) and DeepONet learn mappings between function spaces \cite{li2021fourier,lu2021deeponet,kovachki2023neural}. Adaptive Fourier neural operators (AFNO) mix spectral modes selectively, attention-based models capture long-range dependencies \cite{guibas2022adaptive,geneva2022transformers}, and graph networks learn simulators on particles and unstructured meshes \cite{sanchezgonzalez2020learning,pfaff2021learning}. PDE-Transformer and models pretrained on many families of partial differential equations (PDEs) extend this direction across multiple simulation systems \cite{holzschuh2025pdetransformer,mccabe2024multiple,herde2024poseidon}. The TANTE model learns neural Taylor expansions that adapt the prediction interval to the local temporal complexity \cite{wu2025tante}. For turbulent flows, implicit U-Net enhanced FNOs give stable long-term predictions of three-dimensional large-eddy-simulation fields \cite{li2023iufno}, and a physics-guided spectral-Fourier U-Net operator reconstructs the polymer conformation tensor of viscoelastic turbulence from velocity fields \cite{chen2026physfuno}. These models enlarge the class of updates that a network can represent. In these deterministic surrogates, however, the update is still inferred from past states alone, and the future enters training only as the state to be matched.

A second line of work targets the mismatch between training and deployment histories. Scheduled Sampling replaces observations with model outputs during training, and Professor Forcing aligns teacher-forced and free-running dynamics \cite{bengio2015scheduled,lamb2016professor}. Pushforward training exposes a learned PDE solver to states produced by its own rollouts, and iterative refinement improves long-rollout accuracy \cite{brandstetter2022message,lippe2023pderefiner}. Graph-network and convolutional learned simulators inject training noise for the same purpose \cite{sanchezgonzalez2020learning,pfaff2021learning,stachenfeld2022learned}. Differentiable-solver training shows that the length of the unrolled optimization horizon changes the accuracy of learned corrections \cite{um2020solver,list2022learned}. Learning with privileged information and knowledge distillation, including hint-based guidance of intermediate student representations, transfer training-only information to a deployable predictor \cite{vapnik2009new,hinton2015distilling,lopezpaz2016unifying,karlsson2022timeseries,romero2015fitnets}. For stochastic video generation, Denton and Fergus learn a history-conditioned prior using a target-conditioned posterior and a shared frame predictor \cite{denton2018stochastic}. Latent world models for control share this structure: PlaNet trains a transition prior from past states against a posterior that also sees the current observation \cite{hafner2019planet}, and DreamerV2 introduces KL balancing, which trains the prior toward the posterior faster than it regularizes the posterior toward the prior \cite{hafner2021dreamerv2}.

In recursive flow forecasting, these strategies change which histories the model sees or how its outputs are supervised; the realized future remains a target for predicted states. The realized future can serve a further purpose. Because the next state is available for every training window, it can define what the update should do: the spatial signal that carries the current state to the next one. A deployable model can then be trained to infer that signal from the history alone. Privileged-information and posterior--prior models already use training-only information to shape a deployable predictor. Here we apply this idea to the spatially resolved update signal of a recursive flow forecaster. Unlike the stochastic latent variables of these models, this signal is a deterministic field on the latent state grid that conditions a shared residual transition, and the history-only model that infers it is trained on its own multistep rollouts. We hypothesize that recursive forecasts improve when the update is conditioned on a transition representation that is anchored to the realized future during training, inferable from the history at deployment, and trained on prediction-fed histories.

We test this hypothesis with Future-Aligned Motion Inference for Recursive Physical Forecasting (FAMIR). A convolutional autoencoder maps each physical field to a latent grid and is then frozen. A future-conditioned teacher, which sees the realized next latent state, and a history-conditioned student infer spatial motion tensors on this grid. Both tensors drive one shared residual transition, which thereby learns to read motion that carries the realized change. Posterior--Prior Motion Alignment (PPMA) jointly aligns the two motion representations on observed histories; distilling the student toward a detached teacher instead corresponds to the limit of KL balancing in which the posterior is not pulled toward the prior, and Sec.~\ref{sec:res_ablation} compares the two choices. Prediction-fed multistep objectives train the student on the histories it generates at deployment. At inference the teacher and the realized future are removed; the student, the latent history and the frozen decoder produce the forecast. The contributions are:
\begin{enumerate}
    \item a future-aligned motion formulation in which a future-conditioned teacher and a history-only student each infer a spatial motion tensor that conditions one shared residual latent update, so that information available only during training shapes the signal consumed by the deployed update;
    \item a joint training objective that couples PPMA, observed-history teacher supervision and prediction-fed multistep supervision, together with training-configuration comparisons that identify the roles of prediction-fed training and alignment direction in three flow systems;
    \item a diagnostic analysis of the trained update, combining fixed-state motion interventions, update geometry and the relation between latent and decoded physical errors, which characterizes how the update uses its motion input and shows that latent accuracy predicts decoded accuracy to different degrees in each flow.
\end{enumerate}
Section~\ref{sec:formulation} presents the formulation, Sec.~\ref{sec:results} reports forecasting accuracy, flow statistics, training comparisons and diagnostics, and Sec.~\ref{sec:conclusions} concludes. The governing systems, data and evaluation details are given in the Appendixes.

\section{FAMIR formulation}\label{sec:formulation}

\begin{figure*}[t]
\centering
\includegraphics[width=\textwidth]{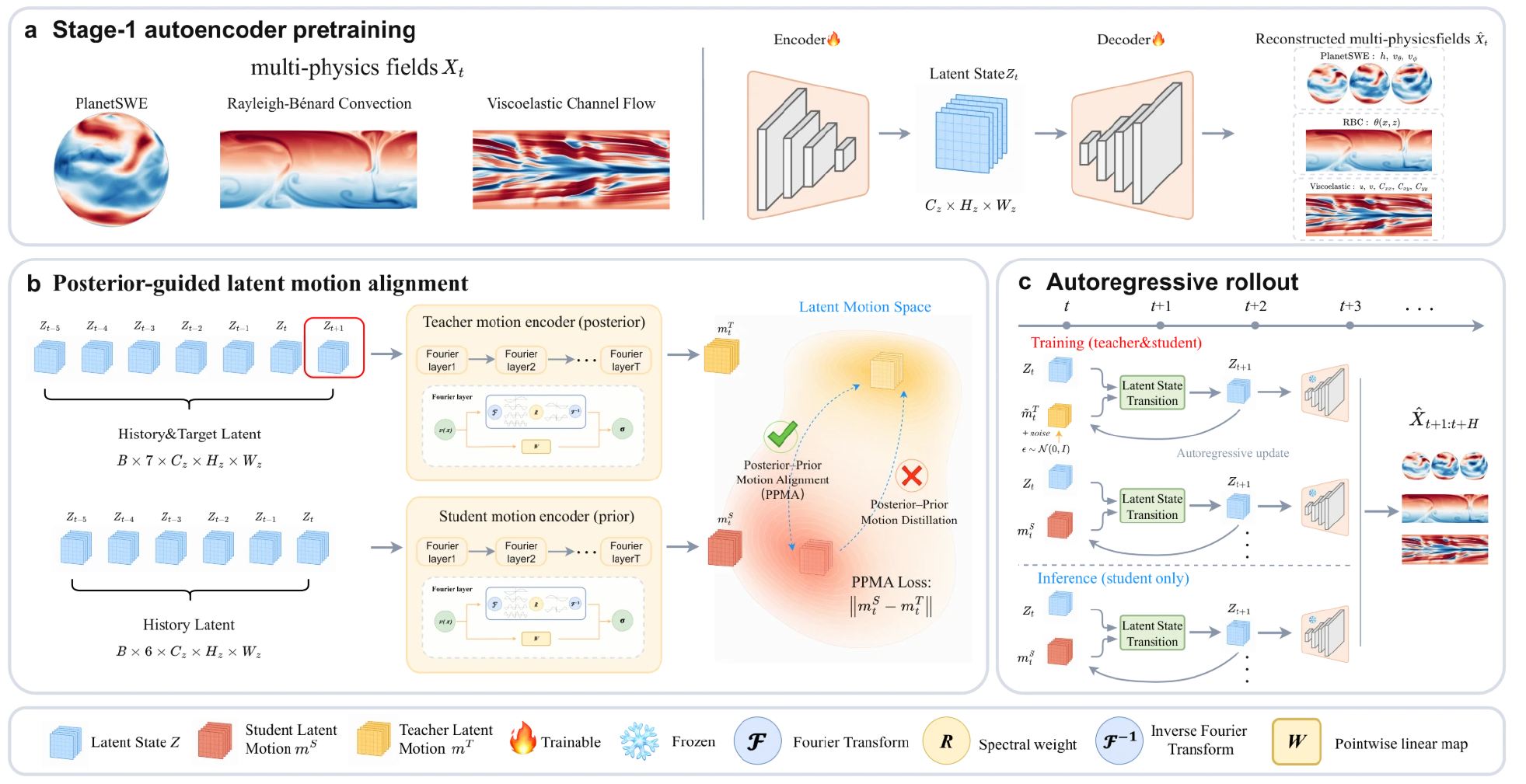}
\caption{FAMIR framework. (a) Stage~1 trains a convolutional autoencoder on the physical fields of each system and then freezes it; the encoder maps each field to a latent state on a spatial grid. (b) Stage~2 motion inference: the teacher sees the latent history and the realized next state $z_{t+1}$, the student sees the history only, and each infers a motion tensor on the latent grid ($m^T_t$, $m^S_t$). PPMA aligns the two motions with gradients to both encoders (check mark); one-way distillation toward a fixed teacher (cross) is the alternative compared as one-way PPMA in Sec.~\ref{sec:res_ablation}. (c) The shared latent transition advances the current state with teacher motion, perturbed by noise during training, or with student motion. The teacher-conditioned update is trained one step at a time on observed histories, the student on its own rollouts; at inference only the student, the transition and the frozen decoder are kept.}
\label{fig:framework}
\end{figure*}

FAMIR separates three questions: how a flow field is represented, what signal drives the change of that representation over one model interval, and how that signal is inferred when only the past is known. Figure~\ref{fig:framework} summarizes the construction. Stage~1 trains and freezes a convolutional autoencoder (Sec.~\ref{sec:latent}). Stage~2 trains, on this latent grid, a history-only student and a future-conditioned teacher that infer motion tensors (Sec.~\ref{sec:motion}) for one shared residual transition (Sec.~\ref{sec:transition}). The objective aligns the two motions and exposes the student to its own rollouts (Sec.~\ref{sec:objective}); deployment keeps only the student (Sec.~\ref{sec:deployment}).

\subsection{Latent representation}\label{sec:latent}

For system $s$, let $X_t^s\in\mathbb{R}^{C_s\times H_s\times W_s}$ be the physical field at time $t$, with consecutive times separated by the model interval of the system (\ref{app:systems}; interval selection in \ref{app:sampling}), and $\mathcal N_s$ the fixed normalization that maps it to the network input. The encoder $E_{\phi_s}$, decoder $D_{\psi_s}$ and six-frame latent history are
\begin{equation}
 x_t^s=\mathcal N_s(X_t^s),\qquad z_t^s=E_{\phi_s}(x_t^s),\qquad
 \mathcal H_t^s=(z_{t-5}^s,\ldots,z_t^s),\qquad
 \widetilde X_t^s=\mathcal N_s^{-1}(D_{\psi_s}(z_t^s)).
 \label{eq:representation}
\end{equation}
The autoencoder is trained by reconstruction on the training split and then frozen, so all neural forecasting models within each system operate in the same latent coordinates. The latent state has 16 channels on a $20\times64$ grid for VF and a $32\times64$ grid for RBC and PlanetSWE. Unlike a global modal vector, this grid retains local adjacency and lets a transition act on localized structures such as conformation bands or plumes. The superscript $s$ is dropped below.

The encoder is a convolutional stack with a width-32 stem, $L_s$ stride-two stages and a $1\times1$ projection to 16 channels; the decoder mirrors it with bilinear upsampling and returns the physical channels through a $3\times3$ output convolution:
\begin{align}
 h^{(0)}_t&=\operatorname{Stem}(x_t), & h^{(\ell+1)}_t&=\operatorname{GELU}\!\left(\operatorname{GN}\!\left(\operatorname{Conv}^{(\ell)}_{3\times3,s=2}(h^{(\ell)}_t)\right)\right),\\
 z_t&=\operatorname{GELU}\!\left(\operatorname{GN}\!\left(\operatorname{Conv}_{1\times1}(h_t^{(L_s)})\right)\right), &
 D_{\psi}(z)&=\operatorname{Conv}_{3\times3}\!\left(\mathcal U_{L_s}\circ\cdots\circ\mathcal U_1(z)\right),
 \label{eq:encoder_decoder}
\end{align}
where GELU is the Gaussian error linear unit, GN denotes group normalization and each $\mathcal U_\ell$ is a bilinear-upsampling convolution--GN--GELU block. VF uses $L_s=4$ stages and RBC and PlanetSWE use $L_s=3$; channel widths and boundary padding are listed in \ref{app:implementation}. POD--Hankel DMD uses its own POD--delay coordinates and does not pass through this autoencoder.

\subsection{Future-aligned motion inference}\label{sec:motion}

The motion tensor $m_t\in\mathbb R^{16\times H_z\times W_z}$, defined on the $H_z\times W_z$ latent grid, is a learned spatial representation of transition information. At every latent-grid site it supplies 16 channels that, together with the local state $z_t$, determine the predicted latent increment over one model interval. We define motion by its role in the transition rather than identify it with a physical velocity or force. Separating motion from the state is the design premise of FAMIR. If a dedicated variable carries how the flow is changing, the transition receives the current state and a separate estimate of its evolution, and that estimate can be supervised with the realized future during training. Two encoders infer it:
\begin{equation}
 m_t^S=P_\omega(\mathcal H_t),\qquad
 m_t^T=Q_\eta(\mathcal H_t,z_{t+1}).
 \label{eq:motion}
\end{equation}
The student $P_\omega$ sees only the history and is the branch used at deployment. The teacher $Q_\eta$ also sees the realized next state $z_{t+1}$, which allows its motion to encode the change that actually occurred. The next state $z_{t+1}$ is available for every training window but never at deployment. Following the prior--posterior terminology of Denton and Fergus \cite{denton2018stochastic}, we also call the student the prior and the teacher the posterior; both are deterministic encoders, and the superscripts $S$ and $T$ mark student and teacher motion.

Both encoders have the same spectral architecture with separate weights. The six (student) or seven (teacher) latent frames are concatenated along channels, two normalized coordinate channels $\xi_x,\xi_y$ are appended, and the result is processed by four Fourier blocks:
\begin{align}
 v^{(0)}&=\operatorname{Lift}([\mathcal H_t,\xi_x,\xi_y]),\\
 v^{(\ell+1)}&=\operatorname{GELU}\!\left(\operatorname{GN}\!\left(W_\ell v^{(\ell)}+
 \mathcal S_\ell(v^{(\ell)})\right)\right),\qquad \ell=0,\ldots,3,\\
 m_t&=\operatorname{Conv}_{3\times3}\!\left(\operatorname{Proj}(v^{(4)})\right),
 \label{eq:fno_block}
\end{align}
where $\operatorname{Lift}$ and $\operatorname{Proj}$ are pointwise convolution--GN--GELU maps and $W_\ell$ is a pointwise projection. The spectral convolution $\mathcal S_\ell$ mixes the $C_{\rm in}$ input channels mode by mode,
\begin{equation*}
\widehat{[\mathcal S_\ell(v)]}_{b,o,k_y,k_x}=\sum_{i=1}^{C_{\rm in}}M_{\ell,o i,k_y,k_x}\,\widehat v_{b,i,k_y,k_x},
\end{equation*}
with $\widehat{\cdot}$ the orthonormal two-dimensional real fast Fourier transform (FFT), $b$ the batch index and $M_\ell$ a learned complex tensor on a truncated set of low modes; higher modes are set to zero before the inverse transform. Through the Fourier blocks, the motion inferred at one site can draw on spatially separated structures of the latent grid.

\subsection{Shared residual transition}\label{sec:transition}

A single transition map is used by both branches:
\begin{align}
 r_t^{(0)}&=\operatorname{GELU}\!\left(\operatorname{GN}\!\left(\operatorname{Conv}_{3\times3}([z_t,m_t])\right)\right),\\
 r_t^{(j+1)}&=\operatorname{GELU}\!\left(r_t^{(j)}+\operatorname{GN}\!\left(\operatorname{Conv}_{3\times3}\,\mathcal C(r_t^{(j)})\right)\right),\quad j=0,1,2,\\
 \widehat{\Delta z}_t&=T_\vartheta(z_t,m_t)=\operatorname{Conv}_{3\times3}(r_t^{(3)}),\\
 F_\vartheta(z_t,m_t)&=z_t+\widehat{\Delta z}_t,\qquad \Delta z_t=z_{t+1}-z_t,
 \label{eq:transition}
\end{align}
where $\mathcal C=\operatorname{GELU}\circ\operatorname{GN}\circ\operatorname{Conv}_{3\times3}$. Only the current state $z_t$ enters the transition; the rest of the history acts through $m_t$. The hat distinguishes the predicted increment $\widehat{\Delta z}_t$ from the observed increment $\Delta z_t$. Sharing $F_\vartheta$ is the route by which the teacher can benefit deployment. Both motion encoders are optimized against next-state targets through the same transition parameters. The student is therefore judged by an update that has also learned from future-informed motion, and its motion is useful to the extent that it carries comparable transition information.

\subsection{Training objective}\label{sec:objective}

Each training window is unrolled along two histories. The reference history $\mathcal H^{\rm ref}_{t+k-1}=(z_{t+k-6},\ldots,z_{t+k-1})$ contains encoded observations only and is shifted by one observed frame per lead $k=1,\ldots,K$. On it, the student and teacher motions are
\begin{equation}
 \bar m^{S}_{k}=P_\omega(\mathcal H^{\rm ref}_{t+k-1}),\qquad \bar m^{T}_{k}=Q_\eta(\mathcal H^{\rm ref}_{t+k-1},z_{t+k}).
 \label{eq:ref_motion}
\end{equation}
PPMA jointly aligns teacher and student motion on the same observed history:
\begin{equation}
 \mathcal L_{\mathrm{PPMA}}=\frac{1}{2\sigma_m^2}\,
 \frac1K\sum_{k=1}^{K}\operatorname{mean}_{b,c,y,x}\!\left[\left(\bar m^{T}_{k}-\bar m^{S}_{k}\right)^2\right],
 \qquad \sigma_m=0.05.
 \label{eq:ppma}
\end{equation}
The square is taken elementwise before averaging over batch, channel and latent grid. Gradients flow into both encoders, so the teacher is also shaped to produce motion that the student can approximate. Before the teacher-conditioned transition, Gaussian noise $\epsilon_k$ is added to $\bar m^T_k$; its standard deviation rises linearly from zero over the first 50 epochs to 0.05 (VF), 0.03 (RBC) or 0.02 (PlanetSWE). PPMA uses the noise-free teacher motion.

The student history $\widehat{\mathcal H}_{t+k-1}$ starts from the same six observations and is then filled with the student's own predictions, $\widehat z_{t+k}=F_\vartheta(\widehat z_{t+k-1},P_\omega(\widehat{\mathcal H}_{t+k-1}))$. This is the input distribution met at deployment. Because gradients are retained through the unrolled rollout, the student learns motion that remains useful when its inputs contain its own errors. With $x_{t+k}$ the normalized target field, the four regression terms are
\begin{align}
 \mathcal L_{\mathrm{lat}}^{S}&=\frac1K\sum_{k=1}^{K}\operatorname{mean}_{b,c,y,x}\!\left[(\widehat z_{t+k}-z_{t+k})^2\right],\\
 \mathcal L_{\mathrm{frm}}^{S}&=\frac1K\sum_{k=1}^{K}\operatorname{mean}_{b,c,y,x}\!\left[(D_{\psi}(\widehat z_{t+k})-x_{t+k})^2\right],\\
 \mathcal L_{\mathrm{prior}}&=\frac1K\sum_{k=1}^{K}\operatorname{mean}_{b,c,y,x}\!\left[(F_\vartheta(z_{t+k-1},\bar m_k^S)-z_{t+k})^2\right],\\
 \mathcal L_{\mathrm{post}}&=\frac1K\sum_{k=1}^{K}\operatorname{mean}_{b,c,y,x}\!\left[(F_\vartheta(z_{t+k-1},\bar m_k^T+\epsilon_k)-z_{t+k})^2\right].
 \label{eq:loss_terms}
\end{align}
The first two supervise the student rollout in latent and decoded normalized-field coordinates; the decoder is frozen but passes gradients to its input. The last two are one-step terms on the reference history for the student and teacher motion, respectively. The objective has three parts: one-step update learning on observed histories for the teacher and the student motion ($\mathcal L_{\rm post}$, $\mathcal L_{\rm prior}$), motion alignment ($\mathcal L_{\rm PPMA}$), and rollout-consistent student forecasting ($\mathcal L^S_{\rm lat}$, $\mathcal L^S_{\rm frm}$). The full objective at epoch $e$ is
\begin{equation}
\begin{split}
 \mathcal L_e={}&a_{\rm pr}(e)\big[\mathcal L_{\rm lat}^{S}+20\mathcal L_{\rm frm}^{S}
 +0.3\mathcal L_{\rm prior}\big]\\
 &+\mathcal L_{\rm post}+a_{\rm ppma}(e)\mathcal L_{\rm PPMA}
 +\mathbf{1}_{\rm SWE}\mathcal L_{\rm seam},
\end{split}
\label{eq:loss}
\end{equation}
where $a_{\rm ppma}(e)$ ramps linearly from zero to one over 50 epochs. For full FAMIR on VF, the weight $a_{\rm pr}(e)$ of the bracketed student terms follows the same ramp; for RBC and PlanetSWE it equals one. The active rollout length $K(e)$ increases from one to five over 100 epochs. The indicator $\mathbf{1}_{\rm SWE}$ selects PlanetSWE, whose seam term penalizes the mismatch between predicted and reference jumps across the periodic longitude boundary.

\subsection{Autonomous forecasting and comparison protocol}\label{sec:deployment}\label{sec:baselines}

At inference, six observed fields are encoded once to initialize $\mathcal H_t$. The student infers motion and advances the last latent state using Eq.~\eqref{eq:transition}; each prediction then replaces the oldest entry in the history. The decoder returns the forecast field $\widehat X_{t+1}=\mathcal N^{-1}(D_{\psi}(\widehat z_{t+1}))$, while recursion proceeds entirely in latent space without observation refresh. The deployed model retains the student motion encoder and shared transition alongside the frozen autoencoder.

The direct latent predictors (latent FNO, latent AFNO and an axial transformer) map the same six-frame latent history to a residual update without a separate motion variable. They share the frozen representation, prediction-fed objective $\mathcal L_{\rm lat}^S+20\mathcal L_{\rm frm}^S$, horizon curriculum and validation-based checkpoint rule; PlanetSWE also uses the common seam term. Unlike FAMIR, the direct predictors receive no observed-history one-step term. POD--Hankel DMD provides a delay-embedded linear reference fitted to training snapshots in POD--delay coordinates.

All neural temporal models of a system also share the data split, optimizer, learning-rate schedule and epoch budget (\ref{app:implementation}). After removal of the teacher, FAMIR deploys 5.82\,M temporal parameters against 5.96--6.03\,M for the direct latent predictors (Table~\ref{tab:cost}); the teacher adds training cost but no deployed parameters. Test histories come from held-out times of the training conditions (VF, RBC) or from held-out trajectories at the training parameters (PlanetSWE). Accuracy is measured on decoded physical fields. The complete-prefix normalized root-mean-square error (NRMSE) pools all leads $1,\ldots,H$ of a forecast and normalizes each field by a fixed range; per-lead NRMSE evaluates a single lead (\ref{app:metrics}). A forecast origin is one initializing six-frame history, and a physical group is one VF condition, RBC Prandtl number or PlanetSWE trajectory.

\section{Results and discussion}\label{sec:results}

\subsection{Recursive forecasting accuracy}\label{sec:res_accuracy}

Table~\ref{tab:accuracy} compares complete-prefix accuracy at ten and twenty steps. FAMIR has the lowest ten-step NRMSE and the highest ten-step peak signal-to-noise ratio (PSNR) and structural similarity (SSIM) in all three systems: 1.003\% for VF, 1.055\% for RBC and 0.620\% for PlanetSWE, which is 29.2\%, 1.8\% and 24.7\% below the strongest direct latent predictor. At twenty steps FAMIR remains the lowest-NRMSE method, with 1.661\%, 1.683\% and 1.036\%, or 16.1\%, 10.9\% and 35.9\% below latent FNO. For VF at twenty steps, POD--Hankel DMD gives the highest PSNR (45.31\,dB, compared with 44.86\,dB for FAMIR), while FAMIR retains the highest SSIM. PSNR averages logarithmic errors per origin and field, so a subset of low-error origins can raise it without lowering the mean NRMSE.

\begin{table}[!htbp]
\centering
\caption{Complete-prefix forecast accuracy at ten and twenty steps. NRMSE is expressed in percent, PSNR in dB, and SSIM is dimensionless. Fields are averaged within origins, origins within physical groups, and groups equally. Bold and underlining mark the best and second-best unrounded scores in each column; shading identifies FAMIR. VF twenty-step scores use the 525 origins shared by all methods.}
\label{tab:accuracy}
\begingroup
\famirtablestyle
\begin{tabular*}{\textwidth}{@{\extracolsep{\fill}}l*{9}{r}@{}}
\multicolumn{10}{@{}l}{\textbf{10-step rollout}} \\[3pt]
\toprule
& \multicolumn{3}{c}{\textbf{VF}} & \multicolumn{3}{c}{\textbf{RBC}} & \multicolumn{3}{c}{\textbf{PlanetSWE}} \\
\cmidrule(lr){2-4}\cmidrule(lr){5-7}\cmidrule(lr){8-10}
Method & NRMSE$\downarrow$ & PSNR$\uparrow$ & SSIM$\uparrow$ & NRMSE$\downarrow$ & PSNR$\uparrow$ & SSIM$\uparrow$ & NRMSE$\downarrow$ & PSNR$\uparrow$ & SSIM$\uparrow$ \\
\midrule
POD--Hankel DMD & 2.274 & 46.36 & 0.8375 & 3.060 & 30.30 & 0.8243 & 2.832 & 31.32 & 0.7697 \\
Latent FNO & \underline{1.416} & 46.72 & \underline{0.9232} & 1.077 & 39.47 & 0.9661 & 0.856 & 41.76 & 0.9644 \\
Latent AFNO & 1.464 & \underline{46.76} & 0.9198 & \underline{1.074} & \underline{39.50} & \underline{0.9663} & 0.911 & 41.21 & 0.9610 \\
Axial transformer & 1.697 & 42.66 & 0.9162 & 1.269 & 38.16 & 0.9571 & \underline{0.823} & \underline{41.91} & \underline{0.9658} \\
\rowcolor{famirtint} \textbf{FAMIR} & \textbf{1.003} & \textbf{48.85} & \textbf{0.9559} & \textbf{1.055} & \textbf{39.64} & \textbf{0.9673} & \textbf{0.620} & \textbf{44.53} & \textbf{0.9817} \\
\bottomrule
\addlinespace[10pt]
\multicolumn{10}{@{}l}{\textbf{20-step rollout}} \\[3pt]
\toprule
& \multicolumn{3}{c}{\textbf{VF}} & \multicolumn{3}{c}{\textbf{RBC}} & \multicolumn{3}{c}{\textbf{PlanetSWE}} \\
\cmidrule(lr){2-4}\cmidrule(lr){5-7}\cmidrule(lr){8-10}
Method & NRMSE$\downarrow$ & PSNR$\uparrow$ & SSIM$\uparrow$ & NRMSE$\downarrow$ & PSNR$\uparrow$ & SSIM$\uparrow$ & NRMSE$\downarrow$ & PSNR$\uparrow$ & SSIM$\uparrow$ \\
\midrule
POD--Hankel DMD & 2.616 & \textbf{45.31} & 0.8218 & 3.270 & 29.73 & 0.8137 & 3.996 & 28.18 & 0.6899 \\
Latent FNO & \underline{1.979} & 43.64 & \underline{0.8904} & \underline{1.890} & \underline{34.61} & \underline{0.9246} & \underline{1.617} & \underline{36.14} & \underline{0.9114} \\
Latent AFNO & 2.085 & 43.77 & 0.8843 & 1.988 & 34.28 & 0.9225 & 2.069 & 33.90 & 0.8908 \\
Axial transformer & 2.374 & 39.68 & 0.8793 & 2.192 & 33.35 & 0.9079 & 1.756 & 35.30 & 0.8981 \\
\rowcolor{famirtint} \textbf{FAMIR} & \textbf{1.661} & \underline{44.86} & \textbf{0.9195} & \textbf{1.683} & \textbf{35.57} & \textbf{0.9334} & \textbf{1.036} & \textbf{40.12} & \textbf{0.9579} \\
\bottomrule
\end{tabular*}
\endgroup

\end{table}

\begin{figure}[!htbp]
\centering
\includegraphics[width=\textwidth]{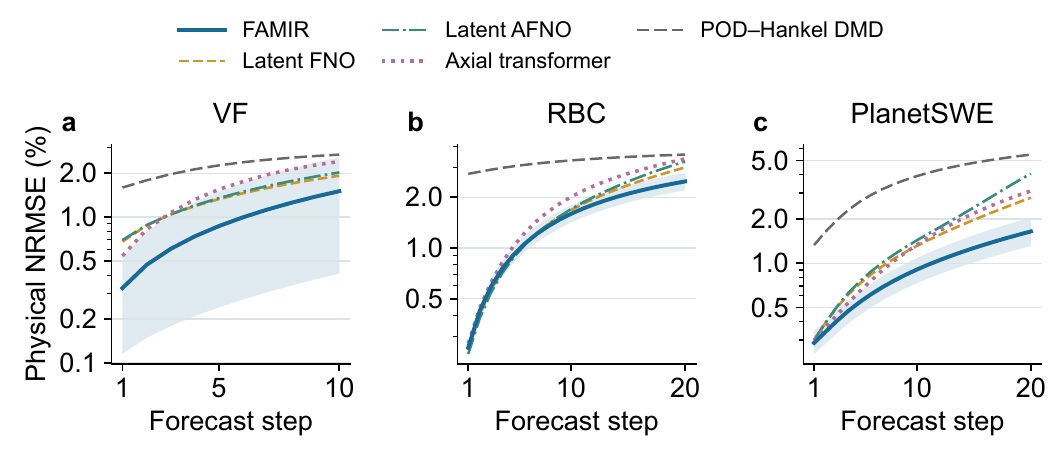}
\caption{Per-lead physical NRMSE during autonomous forecasting. Curves are equal-group means at each lead; the FAMIR band spans the 10th--90th percentiles of the group means, linearly interpolated from three VF, four RBC and six PlanetSWE values. All methods use the same initialized origins within each system. Each point evaluates a single lead, whereas Table~\ref{tab:accuracy} evaluates the complete prefix. VF per-lead errors are shown through ten steps.}
\label{fig:growth}
\end{figure}

The relative gain changes with horizon in different directions across the systems, and Fig.~\ref{fig:growth} shows where it arises. Dividing the FAMIR per-lead NRMSE by the lowest direct-predictor NRMSE at the same lead gives 0.60 at the first VF step and 0.78 at the tenth. In RBC the ratio is 1.09 at the first step, falls below one from the seventh step and reaches 0.83 at the twentieth. In PlanetSWE it decreases from 0.96 to 0.69 at ten steps and 0.59 at twenty. Two forecasting-error patterns emerge. In VF, whose five coupled channels contain sharp conformation gradients, the full FAMIR update is already more accurate than the direct predictors over a single interval, and part of this advantage is lost as errors accumulate. In RBC and PlanetSWE, one-step accuracy is close to that of the direct predictors and the benefit comes from slower error growth under recursion.

\subsection{Flow-field evolution and statistical fidelity}\label{sec:res_fields}

\begin{figure}[!htbp]
\centering
\includegraphics[width=\textwidth,height=.76\textheight,keepaspectratio]{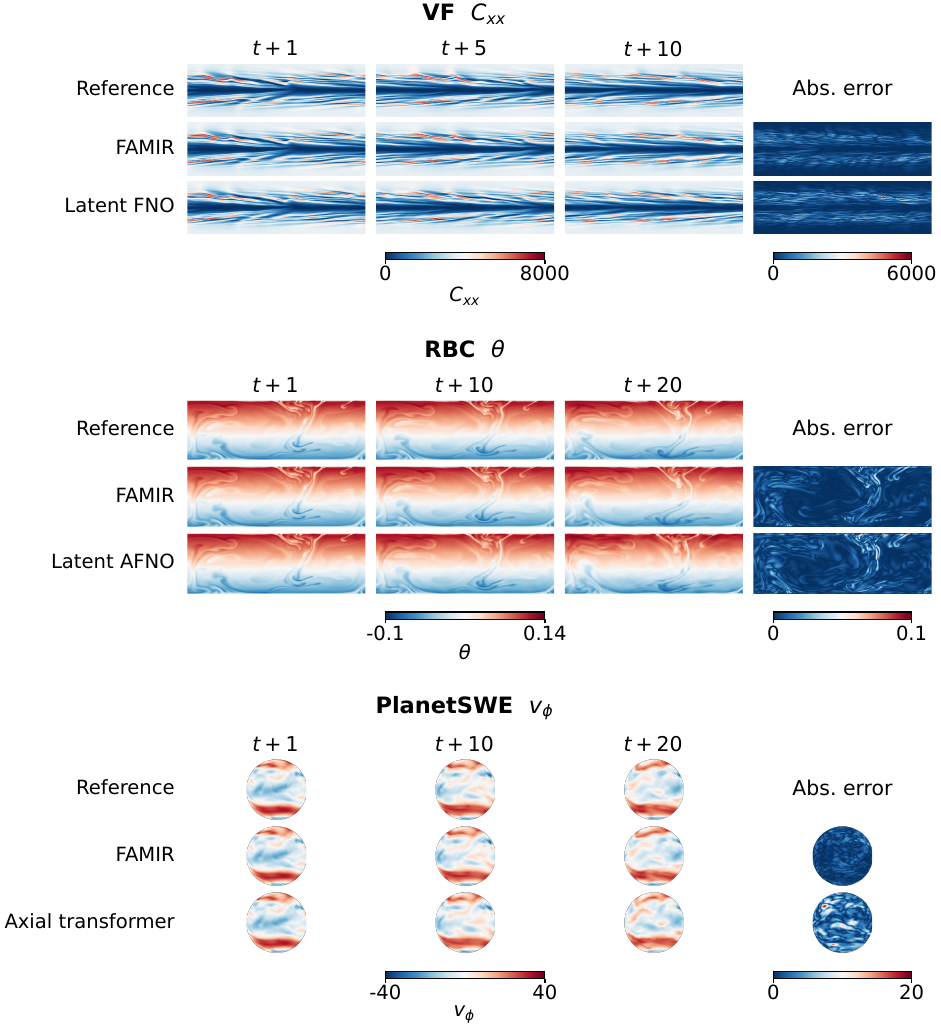}
\caption{Matched field predictions at three leads. Rows show the reference, FAMIR and, as the comparator, the direct latent predictor with the lowest ten-step NRMSE for that system in Table~\ref{tab:accuracy} (at twenty steps, latent FNO is the strongest direct predictor in all systems); the last column shows the absolute error at the last displayed lead (VF $t+10$; RBC and PlanetSWE $t+20$), aligned with each prediction row. VF and RBC panels are drawn with horizontal and vertical axes scaled independently. The VF streamwise conformation component $C_{xx}$ is shown at steps 1, 5 and 10; RBC $\theta$ and PlanetSWE $v_\phi$ at steps 1, 10 and 20. The cases are VF $(EI,Re)=(0.04,1000)$, RBC $(Ra,Pr)=(10^8,1)$ and PlanetSWE test trajectory IC24 (initial condition 24 of The Well), with $EI$ the elasticity number, $Re$ the Reynolds number, $Ra$ the Rayleigh number and $Pr$ the Prandtl number. Model intervals are 1.2 simulation units, 64 lattice-Boltzmann steps and approximately four simulation hours, respectively. Each system uses one field scale and one absolute-error scale for both predictors. RBC panels show the heated plate at the bottom. The $C_{xx}$ scale starts at zero, and the few negative values (below 0.6\% of pixels) are shown at its lower end.}
\label{fig:fields}
\end{figure}

Figure~\ref{fig:fields} places the errors in physical space. In VF, the residuals of both predictors concentrate along the thin conformation bands of $C_{xx}$, where a small displacement of a sharp gradient produces a large pointwise difference. In RBC they follow the edges of rising and falling plumes, and in PlanetSWE they trace displaced zonal-velocity structures. At the displayed terminal leads FAMIR and the comparator reach single-field, single-lead NRMSE values of 4.47\% and 5.77\% in VF, 2.45\% and 3.00\% in RBC, and 1.36\% and 3.22\% in PlanetSWE. On shared color scales the FAMIR residuals are weaker along the same interfaces rather than located elsewhere, indicating more accurate placement of the resolved structures.
\begin{figure}[!htbp]
\centering
\includegraphics[width=.99\textwidth]{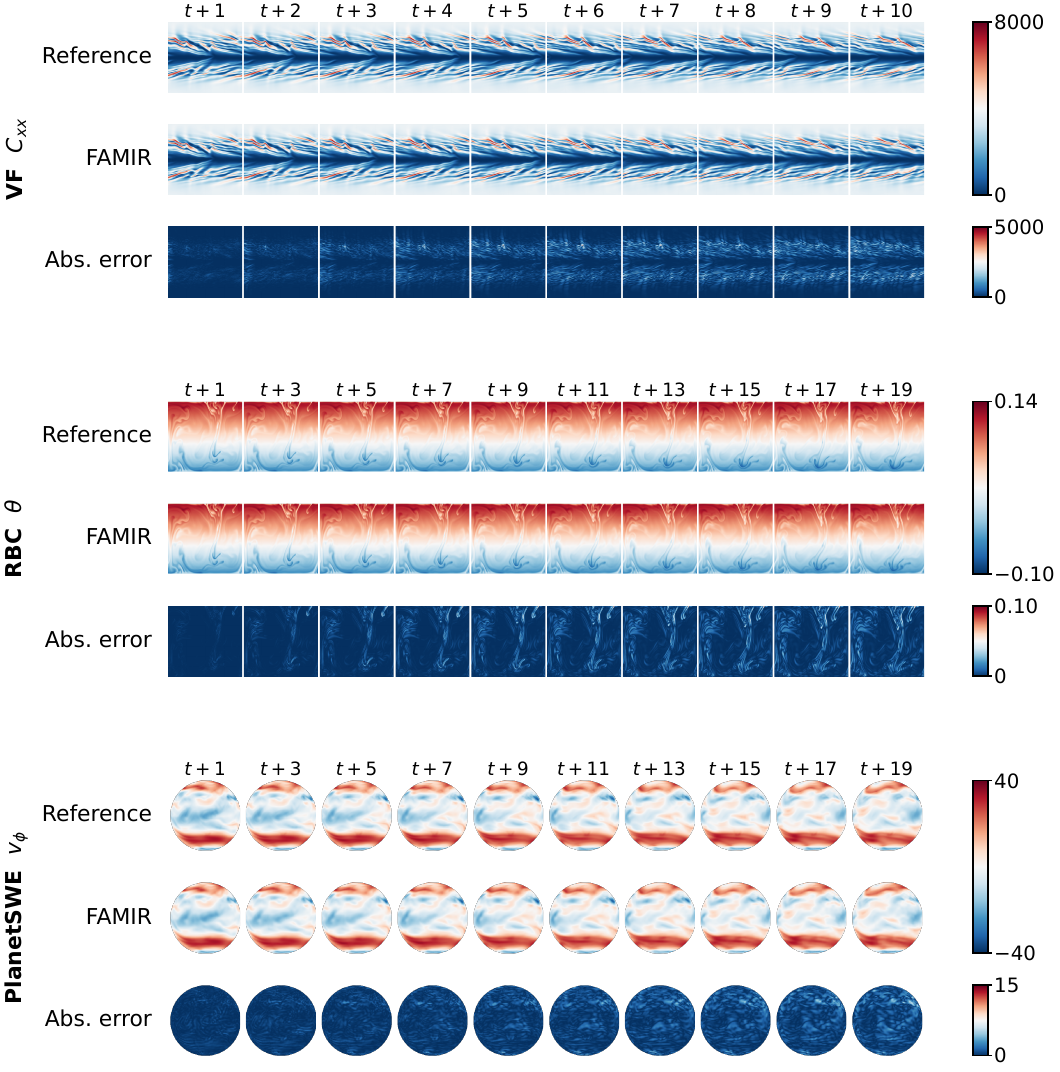}
\caption{Autonomous forecasts at selected leads. Reference, FAMIR prediction and pointwise absolute error are shown for VF at $t+1,\ldots,t+10$ and for RBC and PlanetSWE at the odd leads $t+1,t+3,\ldots,t+19$. Cases and variables match Fig.~\ref{fig:fields}. VF and RBC panels show the full spatial arrays with horizontal and vertical axes scaled independently; PlanetSWE uses a fixed orthographic view. The upper colorbar spans the reference and prediction rows; the lower colorbar scales the error row. Each scale is fixed across the displayed leads. RBC orientation and the $C_{xx}$ scale follow Fig.~\ref{fig:fields} (0.2\% of $C_{xx}$ values lie below zero).}
\label{fig:fullspan}
\end{figure}

Figure~\ref{fig:fullspan} follows the same FAMIR cases at every displayed lead. The conformation bands, plume stems and zonal jets remain in place across the sequence. In VF and RBC the error grows along the bands and plume stems and gradually extends into the surrounding bulk; in PlanetSWE it appears as small filaments over the visible hemisphere. Together with Fig.~\ref{fig:growth}, this indicates that the error grows through gradual deformation and displacement of coherent structures rather than through their loss.
\begin{figure}[!htbp]
\centering
\includegraphics[width=.98\textwidth]{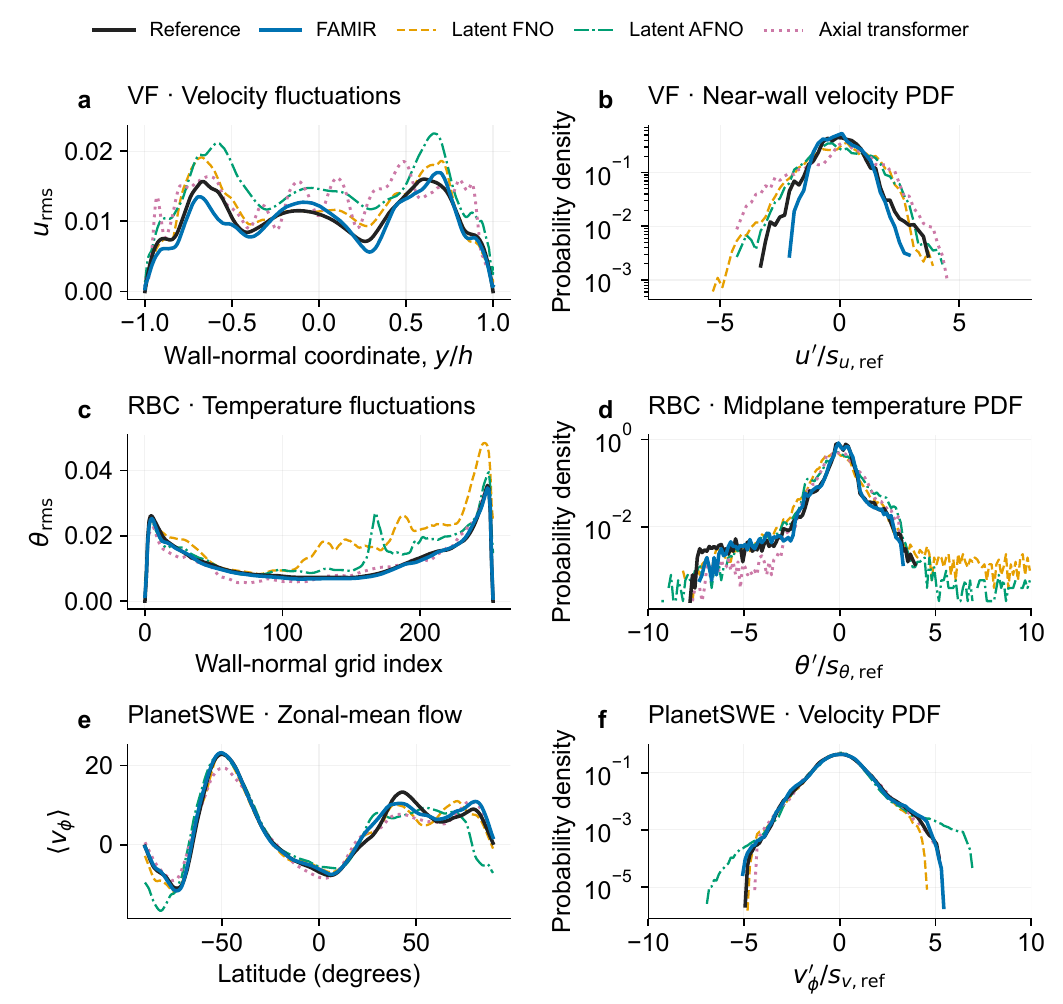}
\caption{Forty-step autonomous flow statistics from one test history per system. VF panels show the streamwise-velocity fluctuation root-mean-square (RMS) and the near-wall velocity probability density function (PDF); RBC panels show the temperature RMS and the midplane temperature PDF; PlanetSWE panels show the zonal-mean velocity and the area-weighted velocity PDF. Curves compare the reference, FAMIR and the three direct latent predictors. The fluctuation RMS and PDF panels are centered on each model's own mean; the PlanetSWE zonal-mean velocity is shown directly. The RBC wall-normal grid index starts at the heated bottom plate. Centering, weights and common reference scales are defined in \ref{app:implementation}.}
\label{fig:statistics}
\end{figure}

Pointwise error does not show whether a forecast retains the fluctuation distribution that governs transport and mixing. Figure~\ref{fig:statistics} therefore compares flow statistics over a forty-step autonomous forecast in each system, twice the longest horizon in Table~\ref{tab:accuracy}, started from the first test history of the first condition or trajectory. Profiles resolve the spatial organization of the flow, and fluctuation PDFs its amplitude distribution. The relative profile and PDF errors of FAMIR are 12.05\% and 19.03\% in VF, against 16.4--37.0\% and 34.3--41.8\% for the direct predictors; 4.09\% and 11.27\% in RBC, against 12.8--58.3\% and 39.0--50.2\%; and 12.37\% and 3.22\% in PlanetSWE, against 16.8--28.5\% and 3.43--10.44\%. The differences lie mainly in profile extrema and PDF tails: several direct predictors overestimate the VF fluctuation peaks near $|y/h|\approx0.65$ and the RBC near-plate peaks and broaden the RBC temperature tails, whereas FAMIR underestimates the lower VF peak by about 14\% (0.0135 against 0.0157). The same ordering holds when the forty-step statistics are repeated for the first test history of each of the first three physical groups (\ref{app:metrics}): FAMIR has the lowest error in 17 of the 18 history--statistic pairs and the lowest three-history mean for every statistic in every system.

\subsection{Effects of the training configuration}\label{sec:res_ablation}

\begin{figure}[!htbp]
\centering
\includegraphics[width=.98\textwidth]{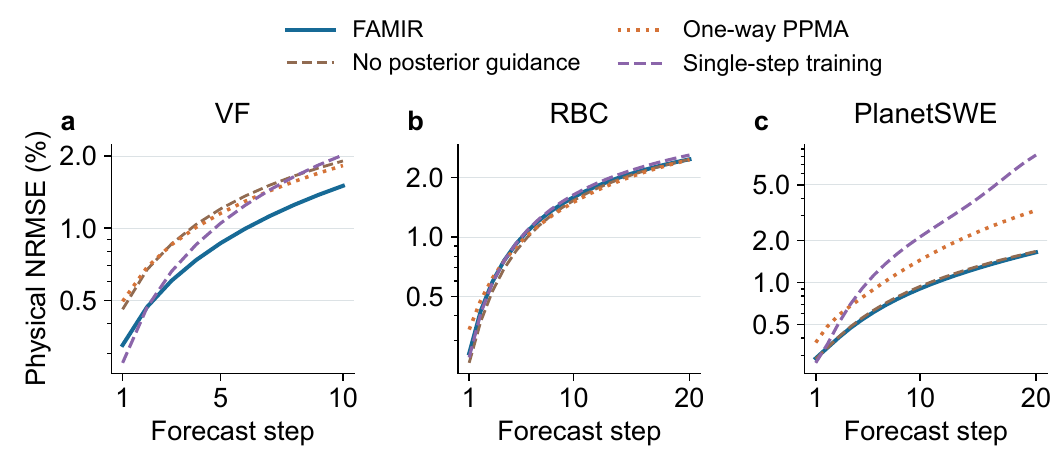}
\caption{Effects of the training configuration on recursive forecast error. Per-lead physical NRMSE is compared for full FAMIR, no posterior guidance, one-way PPMA and single-step training. Curves are equal-group means. VF is shown through ten steps and RBC and PlanetSWE through twenty. Table~\ref{tab:ablation} reports complete-prefix scores.}
\label{fig:ablation}
\end{figure}

The training controls in Table~\ref{tab:ablation} and Fig.~\ref{fig:ablation} are separately trained configurations that share the representation and transition architecture. Scores are complete-prefix NRMSE through ten steps for VF and twenty steps for RBC and PlanetSWE. The no-posterior-guidance configuration removes the teacher encoder $Q_\eta$ and the terms $\mathcal L_{\rm post}$, $\mathcal L_{\rm PPMA}$ and $\mathcal L_{\rm prior}$, leaving the student rollout objectives. One-way PPMA retains $\mathcal L_{\rm post}$ but blocks the teacher-side gradient in $\mathcal L_{\rm PPMA}$. Single-step training keeps the same model family with a one-step objective and no prediction-fed rollout.

\begin{table*}[t]
\centering
\caption{Complete-prefix physical NRMSE (\%) of FAMIR and its training controls, through ten steps for VF and twenty steps for RBC and PlanetSWE. Parentheses give the change relative to FAMIR, computed from unrounded scores; negative values mean lower error.}
\label{tab:ablation}
\famirtablestyle
\begin{tabular*}{\textwidth}{@{\extracolsep{\fill}}lrrr@{}}
\toprule
Configuration & VF (10 steps) & RBC (20 steps) & PlanetSWE (20 steps) \\
\midrule
\rowcolor{famirtint} \textbf{FAMIR} & 1.003 & 1.683 & 1.036 \\
No posterior guidance & 1.333 ($+32.9\%$) & 1.648 ($-2.1\%$) & 1.064 ($+2.7\%$) \\
One-way PPMA & 1.280 ($+27.6\%$) & 1.634 ($-3.0\%$) & 1.874 ($+80.9\%$) \\
Single-step training & 1.284 ($+28.0\%$) & 1.747 ($+3.8\%$) & 3.712 ($+258.3\%$) \\
\bottomrule
\end{tabular*}
\end{table*}

The three flows respond differently to the training configuration (Table~\ref{tab:ablation}). In PlanetSWE, where the FAMIR gain comes from slower accumulation, prediction-fed training is the dominant ingredient: single-step training raises the error by 258.3\%. One-way alignment, in which the teacher motion receives no alignment gradient, raises the error by 80.9\%, whereas removing the teacher altogether raises it by 2.7\%; in this flow, a teacher that is not shaped toward history-inferable motion is worse than no teacher (Sec.~\ref{sec:res_transition}). In VF, where FAMIR gains most over a single interval, every control lies 27.6--32.9\% above full FAMIR; unlike the controls, full FAMIR ramps its student terms over the first 50 epochs in this flow (Sec.~\ref{sec:objective}). In RBC all controls lie within 4\% of full FAMIR, so the twenty-step accuracy of this flow is insensitive to the training configuration.

At the first lead, single-step training, which optimizes exactly this one-step error, is more accurate than full FAMIR in all three flows [0.275\% against 0.326\% in VF, 0.246\% against 0.258\% in RBC and 0.266\% against 0.287\% in PlanetSWE; Fig.~\ref{fig:ablation}], as is the no-posterior-guidance configuration in RBC (0.231\%). The single-step error then grows fastest, and this configuration is the least accurate over the complete prefix. Prediction-fed training thus accepts a larger first-step error in exchange for slower accumulation, most strongly in PlanetSWE.

Even without the teacher, the no-posterior-guidance configuration improves on the best direct predictor in every system: 1.333\% against 1.416\% in VF, 1.648\% against 1.890\% in RBC and 1.064\% against 1.617\% in PlanetSWE. In this configuration the 16-channel motion is an unsupervised intermediate representation, and the model is a two-stage predictor: a spectral history encoder followed by a local convolutional update of the current state. Separating the encoding of the history from the update of the state thus helps in all three flows.

\subsection{Functional role of latent motion}\label{sec:res_motion}

\begin{figure}[!htbp]
\centering
\includegraphics[width=\textwidth]{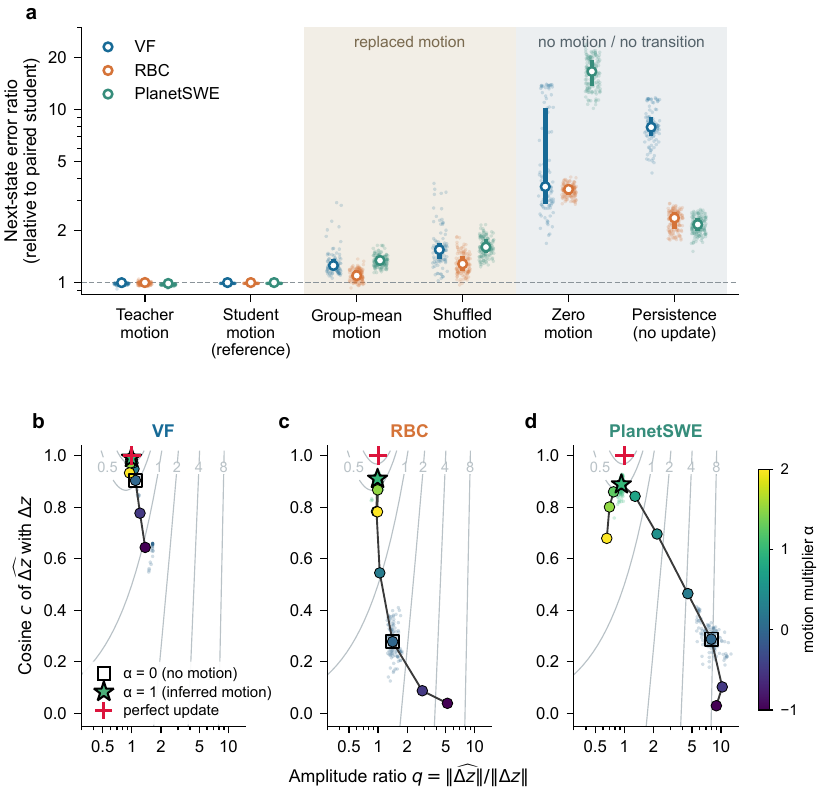}
\caption{Fixed-state interventions on the motion input of the trained FAMIR transition. (a) Next-state error ratio $r_i=\|F_\vartheta(z_i,m)-z_{i+1}\|_2/\|F_\vartheta(z_i,m_i^S)-z_{i+1}\|_2$ when the paired student motion $m_i^S$ is replaced by teacher, leave-one-out group-mean, within-group shuffled or zero motion, and for persistence ($z_i$ in place of $F_\vartheta$); ratios above one indicate a worse forecast of the same state, and the student-motion column equals one by construction. Points show individual origins; open markers and bars show medians and interquartile ranges. Colors denote systems. Shading separates replaced motion (group mean, shuffled) from inputs without a usable motion (zero motion) or without a transition (persistence); the cross in (b--d) marks a perfect update, $q=c=1$. (b--d) Direction and amplitude of the predicted latent increment $\widehat{\Delta z}$ relative to the observed increment $\Delta z$ for scaled motion $m=\alpha m_i^S$. Gray contours give the increment error $\|\widehat{\Delta z}-\Delta z\|_2/\|\Delta z\|_2=(q^2-2qc+1)^{1/2}$, with amplitude ratio $q$ and cosine $c$; the contour at one is the persistence level. Colored markers are medians of $q$ and of $c$ taken separately at each $\alpha$; faint points are individual origins at $\alpha=0$ and $\alpha=1$. Interventions use the fixed-state cohort of 93 VF, 128 RBC and 144 PlanetSWE origins, with state, target and transition held fixed at each.}
\label{fig:controls}
\end{figure}

Figure~\ref{fig:controls} evaluates the response of the trained FAMIR update to its motion input. The motion-replacement tests hold the current latent state, its next-state target and the transition parameters fixed. Shuffling supplies the motion of another origin in the same physical group; mean replacement supplies the group mean excluding the current origin; zero motion supplies a zero tensor to the trained transition. Persistence provides a separate reference that bypasses the transition and returns the current state (\ref{app:motion_diagnostics}). Ratios are formed within each origin relative to its own student motion, so values above one mean that the intervention degrades the forecast of that state.

Figure~\ref{fig:controls}(a) orders these inputs by the information they carry. Substituting teacher motion changes the median error ratio by at most 1.2\%; at a fixed state, teacher motion, although computed from the realized future, gives the trained transition no better update than student motion. Group-mean replacement gives median ratios of 1.26, 1.10 and 1.34 for VF, RBC and PlanetSWE, and within-group shuffling gives 1.55, 1.28 and 1.61. The ordering paired $<$ mean $<$ shuffled holds in all three systems. The motion carries information specific to each state, beyond the motion typical of its physical group, and a mismatched pattern misdirects the update more than the featureless group mean does. RBC shows the smallest penalties for both replacements.

Persistence sets the scale of the change to be predicted. Its ratio is the reciprocal of the increment-relative error, $\|\Delta z_i\|_2/\|\widehat{\Delta z}_i-\Delta z_i\|_2$, and the medians of 7.93, 2.37 and 2.17 correspond to update errors of about 13\%, 42\% and 46\% of the latent change in VF, RBC and PlanetSWE. The VF update captures the largest share of the change, consistent with its first-step advantage in Fig.~\ref{fig:growth}. Zero motion gives median ratios of 3.59, 3.47 and 16.67. In VF this is below persistence: the state input alone already moves the forecast toward the target. In RBC and PlanetSWE it is above persistence; in PlanetSWE its median is nearly eight times that of persistence.

Figure~\ref{fig:controls}(b--d) shows why. Because the transition adds its output to the current state, the error of a forecast is the error of its increment. The increment can be resolved into a direction, the cosine $c$ with the observed increment, and an amplitude, the norm ratio $q$. The two coordinates fix the increment error exactly; a vanishing increment has unit error, so the contour at one marks persistence. Zero motion corresponds to $\alpha=0$. Because the transition never received zero motion in training, this test probes how far the learned update relies on that input. In VF the zero-motion increment already has nearly the right direction and size ($c=0.90$, $q=1.09$); the motion refines it to $c=0.99$, which is why zero motion stays inside the persistence contour. In RBC the zero-motion increment is 41\% too large and poorly aligned with the observed change ($c=0.28$, $q=1.41$). The motion turns it to $c=0.91$, and for $\alpha\ge0.5$ the amplitude ratio remains between 0.96 and 0.99, so in RBC the motion mainly sets the direction of the update. In PlanetSWE the zero-motion increment is misdirected and about eight times too large ($c=0.29$, $q=7.9$); the motion both turns it and removes the excess amplitude ($c=0.89$, $q=0.93$ at $\alpha=1$). The trained transition thus responds to state and motion jointly, and in RBC and PlanetSWE its output without motion is worse than no update.

Along the scaling path, the per-origin error ratio equals the ratio of the increment errors at $\alpha$ and at $\alpha=1$. Its median is lowest at $\alpha=1$ on the evaluated grid in all three systems, as expected for a transition trained on the student's own motion. Reversing the motion gives ratios of 8.29, 12.32 and 19.65 and leaves the RBC and PlanetSWE increments nearly orthogonal to the observed change. The response is nearly symmetric in VF (1.78 at $\alpha=0.5$, 1.69 at $\alpha=1.5$) but asymmetric in RBC (1.51, 1.22) and PlanetSWE (3.64, 1.32). In these two flows weakening the motion costs more than strengthening it, because a smaller $\alpha$ moves the increment toward the poor zero-motion response. The trained update reads the amplitude and sign of the inferred motion as well as its pattern.

\subsection{From latent transitions to physical accuracy}\label{sec:res_transition}

\begin{figure}[!htbp]
\centering
\includegraphics[width=\textwidth]{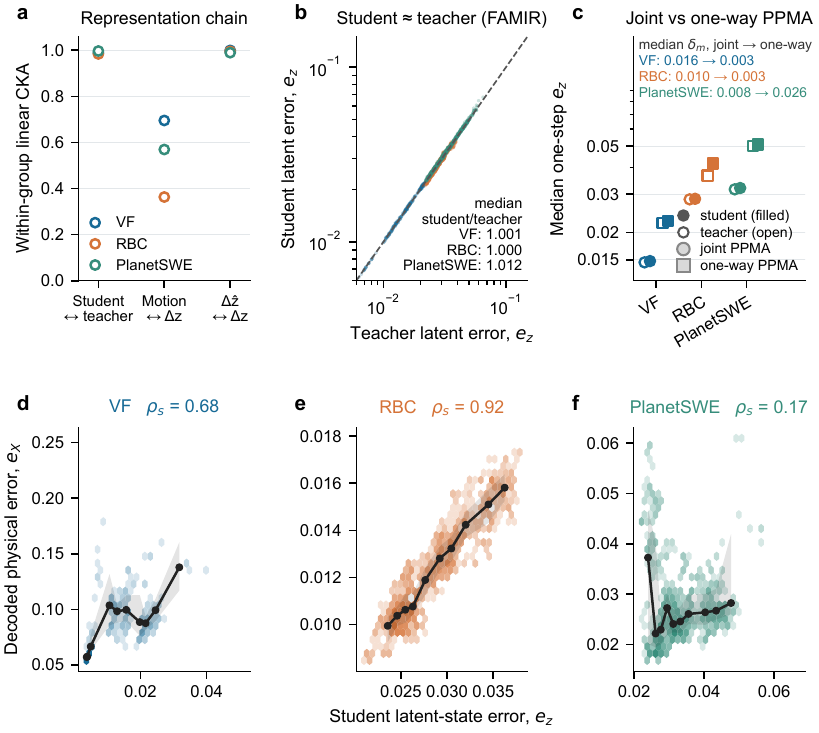}
\caption{From motion to physical error. (a) Within-group linear CKA between student and teacher motion, between student motion and the observed increment $\Delta z$, and between the predicted increment $\widehat{\Delta z}$ and $\Delta z$, on the 93/128/144-origin cohorts of Fig.~\ref{fig:controls}; colors denote systems. (b) Teacher and student one-step relative latent-state error $e_z=\|\widehat z_{t+1}-z_{t+1}\|_2/\|z_{t+1}\|_2$ for individual origins; the dashed line marks equality and the panel lists the median per-origin student-to-teacher ratio. (c) Median one-step $e_z$ of the teacher (open) and student (filled) for full FAMIR with joint PPMA (circles) and for the one-way PPMA configuration (squares); the key gives the median relative motion distance $\delta_m=\|m^S-m^T\|_2/\|m^T\|_2$, joint $\rightarrow$ one-way. (d--f) Decoded physical error $e_X$, the mean of per-channel physical relative $L_2$ errors [Eq.~\eqref{eq:decoded_diagnostic}], against student $e_z$; shading shows the density of origins, and black markers and gray bands give the median and interquartile range of ten equal-count bins; $\rho_s$ is the Spearman correlation. Panels (b--f) use one-step predictions at 525 VF, 1500 RBC and 1362 PlanetSWE origins.}
\label{fig:transition}
\end{figure}

Figure~\ref{fig:transition} follows the update from motion to decoded fields. Panel (a) traces within-group linear centered kernel alignment (CKA) along the chain. Student and teacher motions share nearly the same sample geometry (0.989, 0.984 and 0.997 for VF, RBC and PlanetSWE). Student motion is only partly aligned with the observed increment (0.696, 0.363 and 0.569), yet the predicted increment has nearly the same between-origin geometry as the observed increment (0.9997, 0.9956 and 0.9898). Motion is thus not a copy of the increment but a code read together with the state, consistent with the zero-motion test of Sec.~\ref{sec:res_motion}. Because CKA is invariant to overall rescaling, amplitude errors are measured separately by the increment errors of 13--46\% in Sec.~\ref{sec:res_motion}.

The teacher and student reach nearly the same one-step latent-state error [Fig.~\ref{fig:transition}(b)], with medians of 0.01458 and 0.01478 in VF, 0.02844 and 0.02861 in RBC, and 0.03163 and 0.03207 in PlanetSWE; the per-origin points lie close to the diagonal (median student-to-teacher ratios 1.001, 1.000 and 1.012), and the student medians are within 1.5\% of the teacher medians. The same holds for the increment-relative error on the fixed-state cohorts, with student and teacher medians of 0.1261 and 0.1257, 0.4227 and 0.4163, and 0.4610 and 0.4545. After training, the student infers from the history alone a motion that serves the transition as well as the motion the teacher reads from the realized future, so deploying the student costs no one-step accuracy; the teacher contributes through training (Table~\ref{tab:ablation}).

Panel (c) shows what changes when the teacher motion receives no alignment gradient, in the one-way PPMA configuration of Table~\ref{tab:ablation}. In the one-way configuration both branches are less accurate: the teacher median rises from 0.0146 to 0.0222 in VF, from 0.0284 to 0.0366 in RBC and from 0.0316 to 0.0500 in PlanetSWE, and the student follows. Yet in VF and RBC the one-way student sits closer to its teacher, with the median motion distance $\delta_m$ falling from 0.016 to 0.003 and from 0.010 to 0.003; in PlanetSWE it rises from 0.008 to 0.026. Closer imitation does not yield a better update. In VF and PlanetSWE the lower one-step error of full FAMIR agrees with the 27.6\% and 80.9\% higher complete-prefix error of one-way PPMA in Table~\ref{tab:ablation}. In RBC the one-step difference does not carry over to the twenty-step score, consistent with the small configuration effects in that flow.

Decoding changes the ordering across systems [vertical axes of Fig.~\ref{fig:transition}(d--f)]. The median physical relative error $e_X$ is 0.0885 for VF, 0.0123 for RBC and 0.0258 for PlanetSWE. VF thus has the lowest median latent-state error but the highest decoded error under these definitions; RBC has the lowest decoded error. The two measures weight discrepancies differently. The latent norm is dominated by the largest latent components, whereas $e_X$ normalizes each physical channel by its own energy and averages channels equally, so a weak channel counts as much as a dominant one; VF has five such channels. Physical errors also include the reconstruction contribution of the frozen decoder. Latent-state error magnitudes do not provide a common physical-accuracy scale across the three systems.

Within each system the relation also varies across forecast origins [Fig.~\ref{fig:transition}(d--f)]. RBC exhibits a strong monotonic association ($\rho_s=0.92$), with binned physical errors increasing with latent-state error. VF retains a positive association ($\rho_s=0.68$), but its five-field aggregate shows greater dispersion across origins and nonmonotonic intermediate bins. PlanetSWE shows a weak association ($\rho_s=0.17$): similar latent errors accompany a broad range of decoded errors, and its lowest-latent-error bin even has a higher median physical error than its highest-latent-error bin. Next-state accuracy in latent coordinates and accuracy of the decoded physical fields are distinct levels of forecast quality, and their correspondence depends on the system. Supervising decoded fields through $\mathcal L^S_{\rm frm}$ and evaluating all comparisons on decoded fields (Sec.~\ref{sec:res_accuracy}) accounts for this dependence.

\section{Conclusions}\label{sec:conclusions}

This work asked whether the realized future, which is available for every training window but never at deployment, can shape the update of a recursive flow predictor rather than only supervise its output. FAMIR implements this idea on a frozen spatial latent grid. A future-conditioned teacher and a history-only student infer motion tensors for one shared residual transition, PPMA jointly aligns the two motions, and prediction-fed multistep objectives train the student on its own rollouts. The deployed model keeps only the student and the shared transition, with 5.82\,M temporal parameters. Across viscoelastic channel flow, Rayleigh--B\'enard convection and planetary shallow-water flow, FAMIR gives lower complete-prefix NRMSE than the direct latent predictors and the POD--Hankel DMD reference at ten and twenty steps. Its advantage is largest at the first step in VF, consistent with a more accurate single update, and grows with lead time in RBC and PlanetSWE, consistent with slower error accumulation. The forecast fields keep coherent structures in place, and forty-step flow statistics remain closer to the reference than those of the direct predictors.

The training comparisons and diagnostics show where these gains arise. Separating a history-inferred motion from the state update already improves on the direct predictors in all three flows, and prediction-fed training is decisive in PlanetSWE, where jointly aligning teacher and student is also far more effective than distilling toward a fixed teacher. The trained update relies on the motion specific to each state and on its amplitude; in RBC and PlanetSWE the update without motion is worse than no update. After training, the student matches the one-step accuracy of the future-informed teacher, so dropping the teacher at deployment costs no one-step accuracy. Latent and decoded errors correspond differently in each flow, and such updates are best judged on decoded physical fields.

The present evaluation uses held-out times and trajectories at the training parameters. Natural extensions are forecasting at unseen parameters and in new geometries, three-dimensional flows and longer horizons. The teacher--student structure also suggests a route to uncertainty estimates, since the distance between future-informed and history-inferred motion measures how well the history determines the next update.

\appendix
\counterwithin{figure}{section}
\counterwithin{table}{section}
\renewcommand{\thefigure}{\Alph{section}.\arabic{figure}}
\renewcommand{\thetable}{\Alph{section}.\arabic{table}}
\section{Governing systems and numerical data}\label{app:systems}

\subsection{Viscoelastic channel flow}

The viscoelastic data are direct numerical simulations of two-dimensional elasto-inertial channel flow with the finitely extensible nonlinear elastic model with Peterlin closure (FENE-P), computed with the in-house finite-difference solver of Zhang \textit{et al.}\ \cite{zhang2024minimal}. In nondimensional form,
\begin{align}
 \nabla\!\cdot\boldsymbol u&=0,\\
 \partial_t\boldsymbol u+\boldsymbol u\!\cdot\nabla\boldsymbol u
 &=-\nabla p+\frac{\beta}{Re_s}\nabla^2\boldsymbol u
 +\frac{1-\beta}{Re_s}\nabla\!\cdot\boldsymbol\tau_p+\boldsymbol f,\\
 \partial_t\boldsymbol C+\boldsymbol u\!\cdot\nabla\boldsymbol C
 &=(\nabla\boldsymbol u)\boldsymbol C+\boldsymbol C(\nabla\boldsymbol u)^T
 -\boldsymbol\tau_p,
\end{align}
with $(\nabla\boldsymbol u)_{ij}=\partial_j u_i$, $\boldsymbol\tau_p=Wi_s^{-1}[f_P(\boldsymbol C)\boldsymbol C-\boldsymbol I]$, $f_P(\boldsymbol C)=(L^2-2)/(L^2-\operatorname{tr}\boldsymbol C)$ the Peterlin function for maximum extensibility $L$, and $\boldsymbol f=f(t)\boldsymbol e_x$ a spatially uniform mean pressure gradient adjusted at every step to hold the bulk velocity at unity. All cases use $\beta=0.9$ and $L^2=10000$. The channel is periodic over a streamwise length of 20 and bounded by no-slip walls 2 apart; the stored fields $u,v,C_{xx},C_{xy},C_{yy}$ occupy a $306\times1024$ grid. Time advancement uses a projection method with a streamwise Fourier pressure solution and second-order integration. The three conditions, labeled by elasticity and Reynolds numbers $(EI,Re)=(0.04,1000)$, $(0.05,1000)$ and $(0.06,800)$ in the dataset convention, have solver-convention parameters $(Re_s,Wi_s)=(666.67,26.67)$, $(666.67,33.33)$ and $(533.33,32.00)$, that is, $Re_s=2Re/3$ and $Wi_s=EI\,Re_s$. Each contributes 1000 frames at a model interval of 1.2 simulation units, split contiguously into 700/100/200 training/validation/test frames.

\subsection{Rayleigh--B\'enard convection}

The temperature-driven flow follows the Boussinesq balances
\begin{align}
 \nabla\!\cdot\boldsymbol u&=0,\\
 \partial_t\boldsymbol u+\boldsymbol u\!\cdot\nabla\boldsymbol u
 &=-\rho_0^{-1}\nabla p+\nu\nabla^2\boldsymbol u
 +g\alpha_T(T-T_0)\boldsymbol e_z,\\
 \partial_t T+\boldsymbol u\!\cdot\nabla T&=\kappa\nabla^2T.
\end{align}
Here $Ra=g\alpha_T\Delta T L_z^3/(\nu\kappa)$ and $Pr=\nu/\kappa$. The data use $Ra=10^8$ and $Pr=1,2,5,10$. Thermal lattice Boltzmann method (LBM) simulations are performed on a $512\times8\times256$ grid, periodic horizontally, with no-slip horizontal plates and fixed hot/cold temperatures. After 200000 warm-up steps, each condition supplies 2000 snapshots separated by 64 LBM steps. The learned field is the scaled temperature departure from the conductive profile on the central $x$--$z$ slice; it excludes the solid boundary rows and has shape $254\times512$. Velocity and the full three-dimensional temperature are not predicted. Each condition uses 1400/200/400 frames for training/validation/testing, and time is measured in LBM steps.

\subsection{Planetary shallow-water flow}

PlanetSWE is drawn from the forced rotating shallow-water simulations in The Well \cite{ohana2024well}. In surface-vector notation, the dynamics have the form
\begin{align}
 \partial_t\boldsymbol v+\boldsymbol v\!\cdot\nabla_S\boldsymbol v
 +f\boldsymbol e_r\!\times\boldsymbol v
 &=-g\nabla_S h-\nu_4\nabla_S^4\boldsymbol v,\\
 \partial_t h+\nabla_S\!\cdot[(H_0+h)\boldsymbol v]
 &=F_h-\nu_4\nabla_S^4h,
\end{align}
where $\nabla_S$ denotes surface differentiation, $H_0$ the mean depth, $f$ the latitude-dependent Coriolis parameter, and $F_h$ the prescribed forcing. The source simulations use Dedalus with their published parameters.

The three stored channels are height, velocity toward increasing colatitude $v_\theta$, and eastward velocity $v_\phi$, on a $256\times512$ latitude--longitude grid. Spherical statistics use Gauss--Legendre weights on the stored latitude nodes. Thirty trajectories contain 1008 approximately hourly frames each; indices $0{:}1008{:}4$ give 252 model frames at approximately four-hour intervals. The split assigns 21 complete trajectories to training, three to validation and six to testing, at common source parameters.

\subsection{Temporal sampling}\label{app:sampling}

\begin{figure}[!htbp]
\centering
\includegraphics[width=\textwidth]{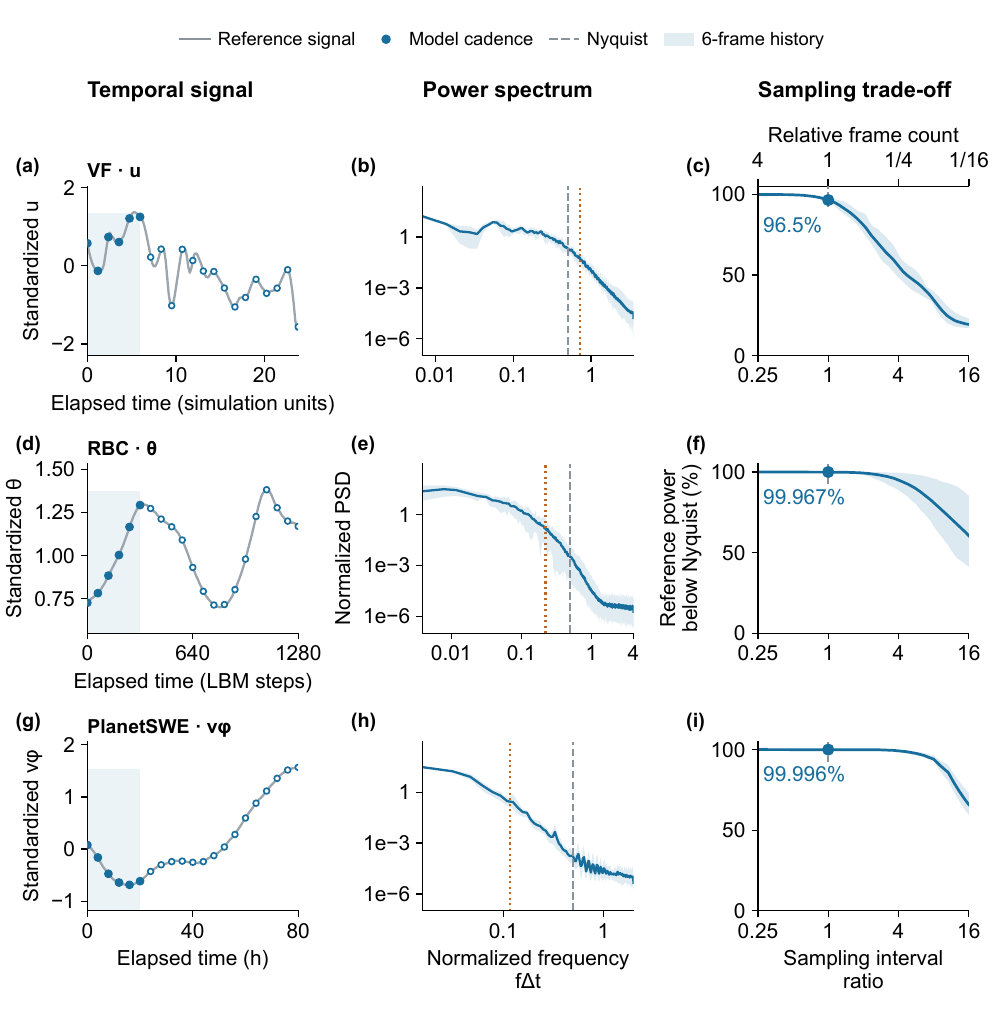}
\caption{Temporal sampling of representative observables. Rows show VF $u$ ($EI=0.04$, $Re=1000$), RBC $\theta$ ($Ra=10^8$, $Pr=1$) and PlanetSWE $v_\phi$. The first column shows dense records and observations at the adopted interval; the six filled observations form the input history. The second column shows probe-averaged, variance-normalized spectra versus $f\Delta t$, with Nyquist and 99\% reference-bandwidth markers. The third column shows reference power below candidate Nyquist frequencies. Adopted intervals are 1.2 simulation units, 64 LBM steps and approximately four simulation hours. Shading gives probe or trajectory ranges.}
\label{fig:sampling}
\end{figure}

The model intervals are assessed with dense records: a VF record seven times denser than the model sequence (7000 frames for the $EI=0.04$ case), a separate $Pr=1$ RBC realization on the same grid sampled every eight LBM steps (16000 frames), and the approximately hourly source frames of the 21 PlanetSWE training trajectories. Twelve fixed probes are used in each system. Spectra use Welch estimation \cite{welch1967spectrum} with Hann windows, 50\% overlap and segment-mean removal, with segments of 1024, 2048 and 256 dense frames; each probe spectrum is normalized by its integrated power before averaging over probes (and, for PlanetSWE, over trajectories). For a reference spectrum $P(f)$ and a candidate interval $\delta$, the power coverage is
\begin{equation}
 \mathcal C(\delta)=
 \frac{\int_0^{1/(2\delta)}P(f)\,df}
 {\int_0^{f_{N,\mathrm{dense}}}P(f)\,df}.
 \label{eq:coverage}
\end{equation}
The adopted intervals cover 96.5\% of the VF velocity power, 99.967\% of the RBC temperature power and 99.996\% of the PlanetSWE zonal-velocity power (Fig.~\ref{fig:sampling}).

\section{Training and evaluation details}\label{app:implementation}

\subsection{Preprocessing, training and model sizes}

VF uses channel-wise min--max scaling to $[0,1]$, and RBC uses min--max scaling to $[-1,1]$. Scaling extrema are fitted to the complete stored arrays before the temporal split: 3000 VF and 8000 RBC frames, including held-out frames. All network weights are fitted on the training splits of \ref{app:systems}. PlanetSWE uses channel-wise means and standard deviations fitted to the 21 training trajectories. Every model within a system shares these transforms, which are inverted before physical evaluation.

\emph{Stage 1.} Autoencoders use AdamW, mixed precision and reconstruction mean-squared error (MSE) for up to 3000 epochs. Batch sizes are 8 for VF and 64 for RBC and PlanetSWE. PlanetSWE adds a longitude-seam penalty of weight 0.1. The selected checkpoints minimize validation reconstruction loss, at epochs 2980, 2980 and 1246, respectively. These frozen autoencoders are shared by all neural temporal models.

\emph{Stage 2.} Training uses AdamW for 300 epochs, weight decay $2\times10^{-4}$, mixed precision and gradient clipping at norm 1. The learning rate rises linearly from $10^{-5}$ to $10^{-4}$ over ten epochs, then follows cosine decay to zero. The rollout horizon grows from one to five over 100 epochs, retaining gradients through predicted histories. Mini-batches contain 4 VF or 16 RBC/PlanetSWE windows. The axial transformer uses mini-batches of 2/16/8 with 4/4/2 accumulation steps, giving effective batches of 8/64/16. Checkpoints minimize the validation student-rollout score $\mathcal L_{\rm lat}^S+20\mathcal L_{\rm frm}^S$; FAMIR selects epochs 295/128/229. Every neural PlanetSWE predictor adds a seam loss of weight one: the squared mismatch between predicted and reference jumps across the longitude boundary, averaged over steps, channels, batch and seam rows.

\emph{Training controls.} Full FAMIR uses the schedules in Eq.~\eqref{eq:loss}. The no-posterior-guidance configuration removes the teacher encoder, teacher-transition loss, PPMA and $0.3\mathcal L_{\rm prior}$. One-way PPMA detaches the teacher only in Eq.~\eqref{eq:ppma}; single-step training fixes the horizon at one. All three VF controls apply their student and prior terms, where present, at full weight from the first epoch. RBC and PlanetSWE use full weight for these terms in every configuration.

\emph{Architectures.} VF has four stride-two encoder stages of widths $64,128,128,192$, with zero padding throughout. RBC and PlanetSWE have three stages of widths $64,128,128$, with circular horizontal/longitude padding and replicated wall-normal/latitude padding. Motion encoders have width 64 and $1\times1$ lift/projection maps. Each spectral block retains five positive and five negative modes on the first axis and sixteen nonnegative real-FFT modes on the second. The shared transition has width 96; its convolutions and the motion head follow the system's encoder padding rule. Latent FNO has four width-68 blocks with modes $(5,16)$; latent AFNO has four width-96 blocks with the same mode budget. The axial transformer has three layers, eight heads and width 288. Each layer applies temporal attention, coordinate-embedded spatial attention and a four-times-expanded GELU feed-forward block with residual normalization. Table~\ref{tab:cost} summarizes deployment sizes.

\begin{table}[!htbp]
\centering
\caption{Deployment parameter counts in millions. The temporal model excludes the teacher. Total size includes the frozen autoencoder: 0.957\,M for VF, 0.502\,M for RBC and 0.503\,M for PlanetSWE.}
\label{tab:cost}
\famirtablestyle
\begin{tabular*}{\textwidth}{@{\extracolsep{\fill}}lrr@{}}
\toprule
Method & Temporal model & With autoencoder \\
\midrule
Latent FNO & 5.959 & 6.461--6.917 \\
Latent AFNO & 6.030 & 6.532--6.987 \\
Axial transformer & 6.007 & 6.508--6.964 \\
\rowcolor{famirtint} \textbf{FAMIR} & 5.820 & 6.322--6.778 \\
\bottomrule
\end{tabular*}
\end{table}

POD--Hankel DMD fits a six-delay linear operator to training snapshots, using one POD bank per physical condition: three for VF, four for RBC and one for PlanetSWE. Validation-selected VF ranks are 699 per field for the first two conditions and 36/133/555/138/140 for $(u,v,C_{xx},C_{xy},C_{yy})$ in the third. RBC retains 1399 temperature modes per condition; PlanetSWE retains 354/4000/4000 modes. Test forecasts apply the fitted operators directly, without refitting or spectral clipping. These ranks approach the number of training snapshots, so the reference is a delay-embedded linear model with little rank reduction. Because each bank belongs to one physical condition, the DMD reference uses the condition label of each test history, which the neural models do not receive.

\begin{figure}[!htbp]
\centering
\includegraphics[width=.90\textwidth]{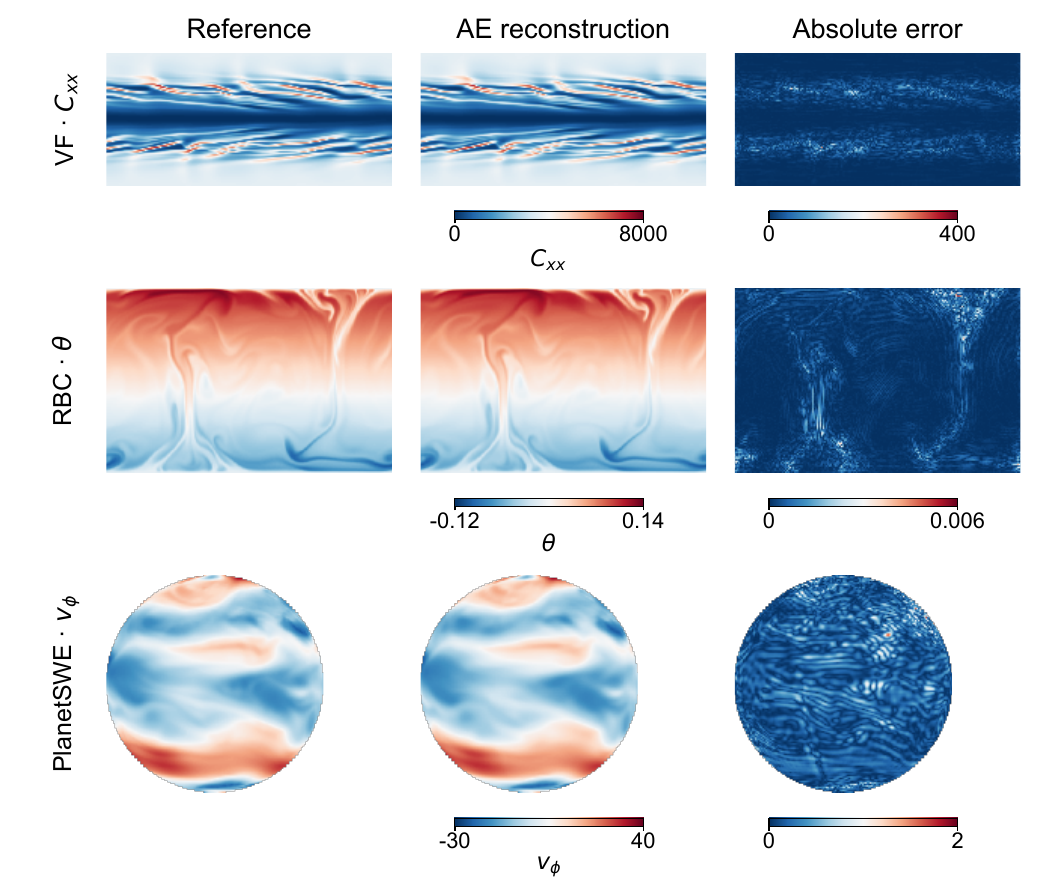}
\caption{Frozen-autoencoder reconstruction. Rows show VF $C_{xx}$, RBC $\theta$ and PlanetSWE $v_\phi$; columns show reference, reconstruction and absolute error. RBC orientation and the $C_{xx}$ scale follow Fig.~\ref{fig:fields}; PlanetSWE uses a fixed orthographic view.}
\label{fig:representation}
\end{figure}

\emph{Spatial reconstruction.} Figure~\ref{fig:representation} checks the frozen representation on the physical fields used to illustrate forecasting. Reference and reconstruction share their physical scale within each system.

\begin{figure}[!htbp]
\centering
\includegraphics[width=.90\textwidth,height=.62\textheight,keepaspectratio]{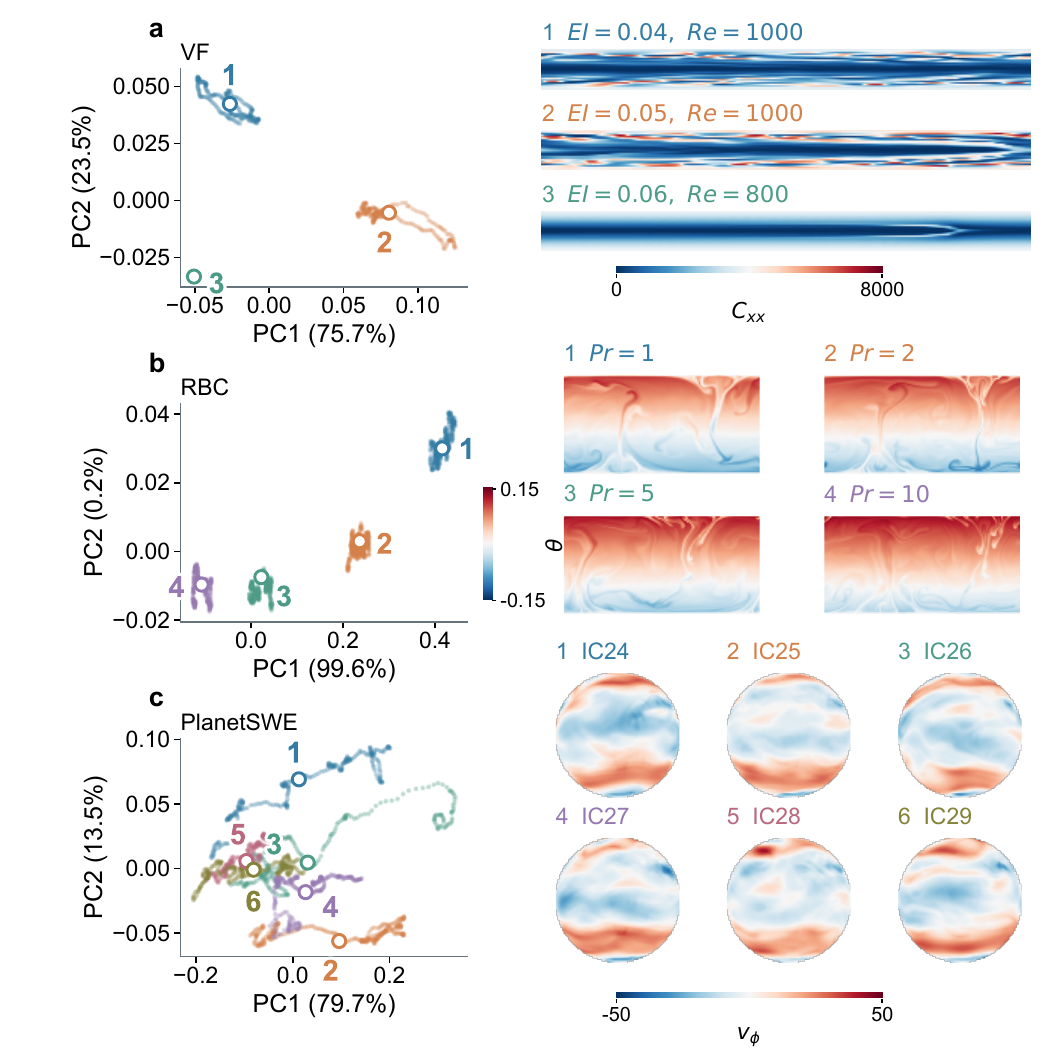}
\caption{Held-out latent summaries paired with observed flow structures. Training-fitted PCA coordinates are colored by the three VF conditions, four RBC Prandtl numbers and six PlanetSWE test trajectories. Open numbered markers identify the within-group representatives; matching numbers and colors identify their reference $C_{xx}$, $\theta$ or $v_\phi$ fields. VF strips retain the 20:2 streamwise-to-wall-normal aspect ratio of the flow domain. A common physical color scale is used within each system; RBC orientation and the $C_{xx}$ scale follow Fig.~\ref{fig:fields}. PC percentages describe variance of the training summaries.}
\label{fig:pca}
\end{figure}

\emph{Latent organization.} Each latent state is summarized by its 16-channel spatial mean. Principal component analysis (PCA), with mean-centering only, is fitted to the training summaries and applied to held-out summaries without refitting. Figure~\ref{fig:pca} shows the PC1--PC2 plane and pairs, within each test condition or trajectory, the summary nearest the group centroid in the first three components with its physical reference field. The three VF conditions occupy separate regions, and the RBC conditions are ordered by Prandtl number along PC1, which carries 99.6\% of the training variance. The six PlanetSWE trajectories, which share physical parameters, largely overlap. The frozen latent space thus separates the VF and RBC conditions, although the condition labels are never given to the autoencoder.

\subsection{Forecast metrics and flow statistics}\label{app:metrics}

For field $c$, origin $i$ and horizon $H$, complete-prefix physical error is
\begin{equation}
 \mathrm{NRMSE}_{i,c}^{(H)}=
 \frac{1}{R_c}\left[\frac{1}{HN_c}\sum_{h=1}^{H}
 \|\widehat X_{i,c}^{(h)}-X_{i,c}^{(h)}\|_2^2\right]^{1/2},
\end{equation}
where $N_c$ is the number of spatial samples. The fixed range $R_c$ uses all stored frames for VF/RBC and the 21 training trajectories for PlanetSWE. Ranges are 1.652, 0.2191, 9016, 4898 and 2666 for VF $(u,v,C_{xx},C_{xy},C_{yy})$, 0.2706 for RBC, and 1906.5, 115.9 and 124.3 for PlanetSWE height/$v_\theta$/$v_\phi$. Per-lead error uses the corresponding single frame. Aggregation averages fields within each origin, origins within each physical group, then groups equally. PSNR is $-20\log_{10}(\mathrm{NRMSE}_{i,c}^{(H)})$ before this aggregation. SSIM uses the same fixed range and a $7\times7$ local window, averaging frames within each prefix before field--origin--group aggregation. Overlapping origins are evaluated as separately initialized forecasts.

The forty-step statistics use six-frame histories with zero-based global indices 800--805 for VF ($EI=0.04$), 1600--1605 for RBC ($Pr=1$) and 6048--6053 for PlanetSWE (IC24). Reference windows are 806--845, 1606--1645 and 6054--6093. These are the first test histories of the first condition (VF, RBC) or test trajectory (PlanetSWE), and each forecast advances autonomously from it. For field $q$, each model forms its own $\bar q(y)=\langle q\rangle_{x,t}$ and $q'=q-\bar q(y)$ over that window. Fluctuation RMS is $\sqrt{\langle q'^2\rangle_{x,t}}$; the PlanetSWE zonal-mean panel displays $\bar q$ directly.

The VF velocity PDF uses the lower-wall grid location nearest $d/h=0.2$, where $d$ is the distance from the wall and $h$ the channel half-height. RBC combines rows 126 and 127 after centering each separately. PlanetSWE pools longitude--time fluctuations with Gauss--Legendre latitude weights. PDF abscissae are $q'/s_{q,\mathrm{ref}}$, where $s_{q,\mathrm{ref}}=\sqrt{\langle(q'_{\mathrm{ref}})^2\rangle_w}$ is the reference fluctuation RMS over the sampled domain. VF uses reference steps 1--60 at the same wall location, centered over that window; RBC pools the separately centered midplane rows over steps 1--40; PlanetSWE uses the full forty-step field after latitude-wise centering. The average uses equal sample weights for VF/RBC and Gauss--Legendre latitude weights for PlanetSWE. All models share this reference scale and common bins: 200 on $[-20,20]$ for VF, and 160 on $[-10,10]$ for RBC/PlanetSWE. Displayed means and samples use steps 1--40. Out-of-range samples are counted separately: 1.07\% for RBC latent FNO and 0.078\% for RBC latent AFNO.

Relative curve error is $[\sum_j w_j(\widehat s_j-s_j)^2/\sum_j w_j s_j^2]^{1/2}$. Profile weights are the stored wall-normal quadrature for VF, equal grid weights for RBC and Gauss--Legendre weights for PlanetSWE. PDF errors use equal-width bins with equal weights.

The three-history comparison of Sec.~\ref{sec:res_fields} uses the first test history of each of the first three physical groups, with zero-based history starts 800, 1800 and 2800 for VF ($EI=0.04$, 0.05, 0.06), 1600, 3600 and 5600 for RBC ($Pr=1$, 2, 5), and 6048, 6300 and 6552 for PlanetSWE (IC24--IC26). Each history is forecast autonomously for forty steps and uses its own reference window, reference scales and the weights defined above.

\subsection{Motion interventions and transition diagnostics}\label{app:motion_diagnostics}

Fixed-state diagnostics use nonoverlapping histories selected by deterministic temporal spacing within the test intervals: 31 origins per VF condition, 32 per RBC condition and 24 per PlanetSWE test trajectory, totaling 93/128/144. Minimum history-start separations are 6/12/6 model frames. At each origin, interventions hold $z_i$, $z_{i+1}$ and the trained transition fixed. Shuffling uses a within-group derangement (seed 20260812); mean replacement excludes the current origin. Zero motion retains the state input, whereas persistence returns $z_i$ directly. Figure~\ref{fig:controls} defines the paired error ratio. Motion scaling uses $\alpha\in\{-1,-0.5,0,0.25,0.5,0.75,1,1.25,1.5,2\}$, with medians and quartiles computed after per-origin ratios.

For Fig.~\ref{fig:transition}(a), tensors are flattened across channels and space and centered across origins within each physical group. Linear CKA \cite{kornblith2019similarity} for centered matrices $A,B$ is
\begin{equation}
 \mathrm{CKA}(A,B)=\frac{\|A^\mathsf T B\|_F^2}
 {\|A^\mathsf T A\|_F\,\|B^\mathsf T B\|_F}.
\end{equation}
The observed increment is $\Delta z_i=z_{i+1}-z_i$, and its relative prediction error is $\|\widehat{\Delta z}_i-\Delta z_i\|_2/\|\Delta z_i\|_2$. Latent-state error instead divides by $\|z_{i+1}\|_2$. After inverse normalization, the decoded physical error is
\begin{equation}
 e_X=\frac{1}{C_s}\sum_{c=1}^{C_s}
 \left[\frac{\sum_{y,x}(\widehat X_{c,y,x}-X_{c,y,x})^2}
 {\max(\sum_{y,x}X_{c,y,x}^2,10^{-12})}\right]^{1/2}.
 \label{eq:decoded_diagnostic}
\end{equation}
Each channel uses its target energy and unweighted spatial sums before channel averaging. This energy normalization differs from the fixed-range NRMSE. Panels (b--f) use one-step forecasts at 525/1500/1362 origins. Correlations pool origins within each system; scatter summaries use ten equal-count bins sorted by latent error, with median coordinates and physical-error interquartile ranges.

\FloatBarrier
\section*{Acknowledgments}
This work was supported by the National Natural Science Foundation of China (12472255, 12202308, 52522609, and 52527901). We thank the contributors to The Well for the PlanetSWE source data.

\section*{Data and code availability}
PlanetSWE source simulations are available through The Well \cite{ohana2024well}. The viscoelastic and Rayleigh--B\'enard data, implementations, checkpoints and evaluation data are available from the corresponding author on reasonable request.

\bibliographystyle{elsarticle-num}
\bibliography{references}

\begin{thebibliography}{10}
\expandafter\ifx\csname url\endcsname\relax
  \def\url#1{\texttt{#1}}\fi
\expandafter\ifx\csname urlprefix\endcsname\relax\def\urlprefix{URL }\fi
\expandafter\ifx\csname href\endcsname\relax
  \def\href#1#2{#2} \def\path#1{#1}\fi

\bibitem{brunton2020machine}
S.~L. Brunton, B.~R. Noack, P.~Koumoutsakos, Machine learning for fluid mechanics, Annual Review of Fluid Mechanics 52 (2020) 477--508.
\newblock \href {https://doi.org/10.1146/annurev-fluid-010719-060214} {\path{doi:10.1146/annurev-fluid-010719-060214}}.

\bibitem{vinuesa2022enhancing}
R.~Vinuesa, S.~L. Brunton, Enhancing computational fluid dynamics with machine learning, Nature Computational Science 2~(6) (2022) 358--366.
\newblock \href {https://doi.org/10.1038/s43588-022-00264-7} {\path{doi:10.1038/s43588-022-00264-7}}.

\bibitem{kochkov2021machine}
D.~Kochkov, J.~A. Smith, A.~Alieva, Q.~Wang, M.~P. Brenner, S.~Hoyer, Machine learning--accelerated computational fluid dynamics, Proceedings of the National Academy of Sciences 118~(21) (2021) e2101784118.
\newblock \href {https://doi.org/10.1073/pnas.2101784118} {\path{doi:10.1073/pnas.2101784118}}.

\bibitem{thuerey2020deep}
N.~Thuerey, K.~Wei{\ss}enow, L.~Prantl, X.~Hu, Deep learning methods for {Reynolds}-averaged {Navier--Stokes} simulations of airfoil flows, AIAA Journal 58~(1) (2020) 25--36.
\newblock \href {https://doi.org/10.2514/1.J058291} {\path{doi:10.2514/1.J058291}}.

\bibitem{berkooz1993proper}
G.~Berkooz, P.~Holmes, J.~L. Lumley, The proper orthogonal decomposition in the analysis of turbulent flows, Annual Review of Fluid Mechanics 25~(1) (1993) 539--575.
\newblock \href {https://doi.org/10.1146/annurev.fl.25.010193.002543} {\path{doi:10.1146/annurev.fl.25.010193.002543}}.

\bibitem{schmid2010dynamic}
P.~J. Schmid, Dynamic mode decomposition of numerical and experimental data, Journal of Fluid Mechanics 656 (2010) 5--28.
\newblock \href {https://doi.org/10.1017/S0022112010001217} {\path{doi:10.1017/S0022112010001217}}.

\bibitem{rowley2017model}
C.~W. Rowley, S.~T.~M. Dawson, Model reduction for flow analysis and control, Annual Review of Fluid Mechanics 49 (2017) 387--417.
\newblock \href {https://doi.org/10.1146/annurev-fluid-010816-060042} {\path{doi:10.1146/annurev-fluid-010816-060042}}.

\bibitem{taira2017modal}
K.~Taira, S.~L. Brunton, S.~T.~M. Dawson, C.~W. Rowley, T.~Colonius, B.~J. McKeon, O.~T. Schmidt, S.~Gordeyev, V.~Theofilis, L.~S. Ukeiley, Modal analysis of fluid flows: An overview, AIAA Journal 55~(12) (2017) 4013--4041.
\newblock \href {https://doi.org/10.2514/1.J056060} {\path{doi:10.2514/1.J056060}}.

\bibitem{mezic2013analysis}
I.~Mezi{\'c}, Analysis of fluid flows via spectral properties of the {Koopman} operator, Annual Review of Fluid Mechanics 45 (2013) 357--378.
\newblock \href {https://doi.org/10.1146/annurev-fluid-011212-140652} {\path{doi:10.1146/annurev-fluid-011212-140652}}.

\bibitem{arbabi2017ergodic}
H.~Arbabi, I.~Mezić, Ergodic theory, dynamic mode decomposition, and computation of spectral properties of the {Koopman} operator, SIAM Journal on Applied Dynamical Systems 16~(4) (2017) 2096--2126.
\newblock \href {https://doi.org/10.1137/17M1125236} {\path{doi:10.1137/17M1125236}}.

\bibitem{brunton2017chaos}
S.~L. Brunton, B.~W. Brunton, J.~L. Proctor, E.~Kaiser, J.~N. Kutz, Chaos as an intermittently forced linear system, Nature Communications 8~(1) (2017) 19.
\newblock \href {https://doi.org/10.1038/s41467-017-00030-8} {\path{doi:10.1038/s41467-017-00030-8}}.

\bibitem{murata2020nonlinear}
T.~Murata, K.~Fukami, K.~Fukagata, Nonlinear mode decomposition with convolutional neural networks for fluid dynamics, Journal of Fluid Mechanics 882 (2020) A13.
\newblock \href {https://doi.org/10.1017/jfm.2019.822} {\path{doi:10.1017/jfm.2019.822}}.

\bibitem{eivazi2020deep}
H.~Eivazi, H.~Veisi, M.~H. Naderi, V.~Esfahanian, Deep neural networks for nonlinear model order reduction of unsteady flows, Physics of Fluids 32~(10) (2020) 105104.
\newblock \href {https://doi.org/10.1063/5.0020526} {\path{doi:10.1063/5.0020526}}.

\bibitem{lusch2018deep}
B.~Lusch, J.~N. Kutz, S.~L. Brunton, Deep learning for universal linear embeddings of nonlinear dynamics, Nature Communications 9~(1) (2018) 4950.
\newblock \href {https://doi.org/10.1038/s41467-018-07210-0} {\path{doi:10.1038/s41467-018-07210-0}}.

\bibitem{otto2019linearly}
S.~E. Otto, C.~W. Rowley, Linearly recurrent autoencoder networks for learning dynamics, SIAM Journal on Applied Dynamical Systems 18~(1) (2019) 558--593.
\newblock \href {https://doi.org/10.1137/18M1177846} {\path{doi:10.1137/18M1177846}}.

\bibitem{brunton2016discovering}
S.~L. Brunton, J.~L. Proctor, J.~N. Kutz, Discovering governing equations from data by sparse identification of nonlinear dynamical systems, Proceedings of the National Academy of Sciences 113~(15) (2016) 3932--3937.
\newblock \href {https://doi.org/10.1073/pnas.1517384113} {\path{doi:10.1073/pnas.1517384113}}.

\bibitem{fukami2021sindy}
K.~Fukami, T.~Murata, K.~Zhang, K.~Fukagata, Sparse identification of nonlinear dynamics with low-dimensionalized flow representations, Journal of Fluid Mechanics 926 (2021) A10.
\newblock \href {https://doi.org/10.1017/jfm.2021.697} {\path{doi:10.1017/jfm.2021.697}}.

\bibitem{wiewel2019latent}
S.~Wiewel, M.~Becher, N.~Thuerey, Latent space physics: Towards learning the temporal evolution of fluid flow, Computer Graphics Forum 38~(2) (2019) 71--82.
\newblock \href {https://doi.org/10.1111/cgf.13620} {\path{doi:10.1111/cgf.13620}}.

\bibitem{vlachas2022multiscale}
P.~R. Vlachas, G.~Arampatzis, C.~Uhler, P.~Koumoutsakos, Multiscale simulations of complex systems by learning their effective dynamics, Nature Machine Intelligence 4~(4) (2022) 359--366.
\newblock \href {https://doi.org/10.1038/s42256-022-00464-w} {\path{doi:10.1038/s42256-022-00464-w}}.

\bibitem{srinivasan2019predictions}
P.~A. Srinivasan, L.~Guastoni, H.~Azizpour, P.~Schlatter, R.~Vinuesa, Predictions of turbulent shear flows using deep neural networks, Physical Review Fluids 4~(5) (2019) 054603.
\newblock \href {https://doi.org/10.1103/PhysRevFluids.4.054603} {\path{doi:10.1103/PhysRevFluids.4.054603}}.

\bibitem{nakamura2021cnnlstm}
T.~Nakamura, K.~Fukami, K.~Hasegawa, Y.~Nabae, K.~Fukagata, Convolutional neural network and long short-term memory based reduced order surrogate for minimal turbulent channel flow, Physics of Fluids 33~(2) (2021) 025116.
\newblock \href {https://doi.org/10.1063/5.0039845} {\path{doi:10.1063/5.0039845}}.

\bibitem{solerarico2024beta}
A.~Solera-Rico, C.~S. Vila, M.~Gómez-López, Y.~Wang, A.~Almashjary, S.~T.~M. Dawson, R.~Vinuesa, {$\beta$}-variational autoencoders and transformers for reduced-order modelling of fluid flows, Nature Communications 15~(1) (2024) 1361.
\newblock \href {https://doi.org/10.1038/s41467-024-45578-4} {\path{doi:10.1038/s41467-024-45578-4}}.

\bibitem{kontolati2024learning}
K.~Kontolati, S.~Goswami, G.~E. Karniadakis, M.~D. Shields, Learning nonlinear operators in latent spaces for real-time predictions of complex dynamics in physical systems, Nature Communications 15~(1) (2024) 5101.
\newblock \href {https://doi.org/10.1038/s41467-024-49411-w} {\path{doi:10.1038/s41467-024-49411-w}}.

\bibitem{li2021fourier}
Z.~Li, N.~Kovachki, K.~Azizzadenesheli, B.~Liu, K.~Bhattacharya, A.~Stuart, A.~Anandkumar, \href{https://openreview.net/forum?id=c8P9NQVtmnO}{{Fourier} neural operator for parametric partial differential equations}, in: International Conference on Learning Representations, 2021.
\newline\urlprefix\url{https://openreview.net/forum?id=c8P9NQVtmnO}

\bibitem{lu2021deeponet}
L.~Lu, P.~Jin, G.~Pang, Z.~Zhang, G.~E. Karniadakis, Learning nonlinear operators via {DeepONet} based on the universal approximation theorem of operators, Nature Machine Intelligence 3~(3) (2021) 218--229.
\newblock \href {https://doi.org/10.1038/s42256-021-00302-5} {\path{doi:10.1038/s42256-021-00302-5}}.

\bibitem{kovachki2023neural}
N.~Kovachki, Z.~Li, B.~Liu, K.~Azizzadenesheli, K.~Bhattacharya, A.~Stuart, A.~Anandkumar, \href{https://www.jmlr.org/papers/v24/21-1524.html}{Neural operator: Learning maps between function spaces with applications to {PDEs}}, Journal of Machine Learning Research 24~(89) (2023) 1--97.
\newline\urlprefix\url{https://www.jmlr.org/papers/v24/21-1524.html}

\bibitem{guibas2022adaptive}
J.~Guibas, M.~Mardani, Z.~Li, A.~Tao, A.~Anandkumar, B.~Catanzaro, \href{https://openreview.net/forum?id=EXHG-A3jlM}{Efficient token mixing for transformers via adaptive {Fourier} neural operators}, in: International Conference on Learning Representations, 2022.
\newline\urlprefix\url{https://openreview.net/forum?id=EXHG-A3jlM}

\bibitem{geneva2022transformers}
N.~Geneva, N.~Zabaras, Transformers for modeling physical systems, Neural Networks 146 (2022) 272--289.
\newblock \href {https://doi.org/10.1016/j.neunet.2021.11.022} {\path{doi:10.1016/j.neunet.2021.11.022}}.

\bibitem{sanchezgonzalez2020learning}
A.~Sanchez-Gonzalez, J.~Godwin, T.~Pfaff, R.~Ying, J.~Leskovec, P.~W. Battaglia, Learning to simulate complex physics with graph networks, in: Proceedings of the 37th International Conference on Machine Learning, Vol. 119 of Proceedings of Machine Learning Research, 2020, pp. 8459--8468.

\bibitem{pfaff2021learning}
T.~Pfaff, M.~Fortunato, A.~Sanchez-Gonzalez, P.~W. Battaglia, Learning mesh-based simulation with graph networks, in: International Conference on Learning Representations, 2021.

\bibitem{holzschuh2025pdetransformer}
B.~Holzschuh, Q.~Liu, G.~Kohl, N.~Thuerey, \href{https://proceedings.mlr.press/v267/holzschuh25a.html}{{PDE-Transformer}: Efficient and versatile transformers for physics simulations}, in: Proceedings of the 42nd International Conference on Machine Learning, Vol. 267 of Proceedings of Machine Learning Research, PMLR, 2025, pp. 23562--23602.
\newline\urlprefix\url{https://proceedings.mlr.press/v267/holzschuh25a.html}

\bibitem{mccabe2024multiple}
M.~McCabe, B.~R{\'e}galdo-Saint~Blancard, L.~H. Parker, R.~Ohana, M.~Cranmer, A.~Bietti, M.~Eickenberg, S.~Golkar, G.~Krawezik, F.~Lanusse, M.~Pettee, T.~Tesileanu, K.~Cho, S.~Ho, Multiple physics pretraining for spatiotemporal surrogate models, in: Advances in Neural Information Processing Systems, Vol.~37, 2024.
\newblock \href {https://doi.org/10.52202/079017-3791} {\path{doi:10.52202/079017-3791}}.

\bibitem{herde2024poseidon}
M.~Herde, B.~Raoni{\'c}, T.~Rohner, R.~K{\"a}ppeli, R.~Molinaro, E.~de~B{\'e}zenac, S.~Mishra, Poseidon: Efficient foundation models for {PDEs}, in: Advances in Neural Information Processing Systems, Vol.~37, 2024.
\newblock \href {https://doi.org/10.52202/079017-2311} {\path{doi:10.52202/079017-2311}}.

\bibitem{wu2025tante}
Z.~Wu, S.~Wang, S.~Zhang, S.~He, M.~Zhu, A.~Jiao, L.~Lu, D.~van Dijk, \href{https://doi.org/10.1016/j.jcp.2026.115041}{{TANTE}: Time-adaptive operator learning via neural {Taylor} expansion}, Journal of Computational Physics 562 (2026) 115041.
\newblock \href {https://doi.org/10.1016/j.jcp.2026.115041} {\path{doi:10.1016/j.jcp.2026.115041}}.
\newline\urlprefix\url{https://doi.org/10.1016/j.jcp.2026.115041}

\bibitem{li2023iufno}
Z.~Li, W.~Peng, Z.~Yuan, J.~Wang, Long-term predictions of turbulence by implicit {U-Net} enhanced {Fourier} neural operator, Physics of Fluids 35~(7) (2023) 075145.
\newblock \href {https://doi.org/10.1063/5.0158830} {\path{doi:10.1063/5.0158830}}.

\bibitem{chen2026physfuno}
Y.~Chen, Y.~Wang, H.~Cheng, H.~Zhang, Y.~Li, W.~Zhang, X.~Li, F.~Li, A physics-guided spectral-{Fourier} {U-Net} operator: From velocity field to hidden tensor field in viscoelastic turbulence, Physics of Fluids 38~(7) (2026) 073103.
\newblock \href {https://doi.org/10.1063/5.0333863} {\path{doi:10.1063/5.0333863}}.

\bibitem{bengio2015scheduled}
S.~Bengio, O.~Vinyals, N.~Jaitly, N.~Shazeer, \href{https://proceedings.neurips.cc/paper/2015/hash/e995f98d56967d946471af29d7bf99f1-Abstract.html}{Scheduled sampling for sequence prediction with recurrent neural networks}, in: Advances in Neural Information Processing Systems, Vol.~28, 2015.
\newline\urlprefix\url{https://proceedings.neurips.cc/paper/2015/hash/e995f98d56967d946471af29d7bf99f1-Abstract.html}

\bibitem{lamb2016professor}
A.~M. Lamb, A.~Goyal, Y.~Zhang, S.~Zhang, A.~C. Courville, Y.~Bengio, \href{https://papers.neurips.cc/paper/2016/hash/16026d60ff9b54410b3435b403afd226-Abstract.html}{Professor forcing: A new algorithm for training recurrent networks}, in: Advances in Neural Information Processing Systems, Vol.~29, 2016.
\newline\urlprefix\url{https://papers.neurips.cc/paper/2016/hash/16026d60ff9b54410b3435b403afd226-Abstract.html}

\bibitem{brandstetter2022message}
J.~Brandstetter, D.~Worrall, M.~Welling, \href{https://openreview.net/forum?id=vSix3HPYKSU}{Message passing neural {PDE} solvers}, in: International Conference on Learning Representations, 2022.
\newline\urlprefix\url{https://openreview.net/forum?id=vSix3HPYKSU}

\bibitem{lippe2023pderefiner}
P.~Lippe, B.~S. Veeling, P.~Perdikaris, R.~E. Turner, J.~Brandstetter, \href{https://openreview.net/forum?id=Qv6468llWS}{{PDE-Refiner}: Achieving accurate long rollouts with neural {PDE} solvers}, in: Advances in Neural Information Processing Systems, Vol.~36, 2023, pp. 67398--67433.
\newline\urlprefix\url{https://openreview.net/forum?id=Qv6468llWS}

\bibitem{stachenfeld2022learned}
K.~Stachenfeld, D.~B. Fielding, D.~Kochkov, M.~Cranmer, T.~Pfaff, J.~Godwin, C.~Cui, S.~Ho, P.~Battaglia, A.~Sanchez-Gonzalez, Learned coarse models for efficient turbulence simulation, in: International Conference on Learning Representations, 2022.

\bibitem{um2020solver}
K.~Um, R.~Brand, Y.~R. Fei, P.~Holl, N.~Thuerey, Solver-in-the-loop: Learning from differentiable physics to interact with iterative {PDE}-solvers, in: Advances in Neural Information Processing Systems, Vol.~33, 2020.

\bibitem{list2022learned}
B.~List, L.-W. Chen, N.~Thuerey, Learned turbulence modelling with differentiable fluid solvers: physics-based loss functions and optimisation horizons, Journal of Fluid Mechanics 949 (2022) A25.
\newblock \href {https://doi.org/10.1017/jfm.2022.738} {\path{doi:10.1017/jfm.2022.738}}.

\bibitem{vapnik2009new}
V.~Vapnik, A.~Vashist, A new learning paradigm: Learning using privileged information, Neural Networks 22~(5-6) (2009) 544--557.
\newblock \href {https://doi.org/10.1016/j.neunet.2009.06.042} {\path{doi:10.1016/j.neunet.2009.06.042}}.

\bibitem{hinton2015distilling}
G.~Hinton, O.~Vinyals, J.~Dean, \href{https://arxiv.org/abs/1503.02531}{Distilling the knowledge in a neural network}, arXiv preprint arXiv:1503.02531 (2015).
\newline\urlprefix\url{https://arxiv.org/abs/1503.02531}

\bibitem{lopezpaz2016unifying}
D.~Lopez-Paz, L.~Bottou, B.~Sch{\"o}lkopf, V.~Vapnik, Unifying distillation and privileged information, in: International Conference on Learning Representations, 2016.

\bibitem{karlsson2022timeseries}
R.~K. Karlsson, M.~Willbo, Z.~M. Hussain, R.~G. Krishnan, D.~Sontag, F.~Johansson, \href{https://proceedings.mlr.press/v151/k-a-karlsson22a.html}{Using time-series privileged information for provably efficient learning of prediction models}, in: Proceedings of the 25th International Conference on Artificial Intelligence and Statistics, Vol. 151 of Proceedings of Machine Learning Research, PMLR, 2022, pp. 5459--5484.
\newline\urlprefix\url{https://proceedings.mlr.press/v151/k-a-karlsson22a.html}

\bibitem{romero2015fitnets}
A.~Romero, N.~Ballas, S.~E. Kahou, A.~Chassang, C.~Gatta, Y.~Bengio, \href{https://arxiv.org/abs/1412.6550}{{FitNets}: Hints for thin deep nets}, in: International Conference on Learning Representations, 2015.
\newline\urlprefix\url{https://arxiv.org/abs/1412.6550}

\bibitem{denton2018stochastic}
E.~Denton, R.~Fergus, \href{https://proceedings.mlr.press/v80/denton18a.html}{Stochastic video generation with a learned prior}, in: Proceedings of the 35th International Conference on Machine Learning, Vol.~80 of Proceedings of Machine Learning Research, PMLR, 2018, pp. 1174--1183.
\newline\urlprefix\url{https://proceedings.mlr.press/v80/denton18a.html}

\bibitem{hafner2019planet}
D.~Hafner, T.~Lillicrap, I.~Fischer, R.~Villegas, D.~Ha, H.~Lee, J.~Davidson, Learning latent dynamics for planning from pixels, in: Proceedings of the 36th International Conference on Machine Learning, Vol.~97 of Proceedings of Machine Learning Research, 2019, pp. 2555--2565.

\bibitem{hafner2021dreamerv2}
D.~Hafner, T.~Lillicrap, M.~Norouzi, J.~Ba, Mastering {Atari} with discrete world models, in: International Conference on Learning Representations, 2021.

\bibitem{zhang2024minimal}
H.~Zhang, H.~Cheng, S.~Wang, W.~Zhang, X.~Li, F.~Li, The minimal flow unit and origin of two-dimensional elasto-inertial turbulence, Journal of Fluid Mechanics 999 (2024) A82.
\newblock \href {https://doi.org/10.1017/jfm.2024.977} {\path{doi:10.1017/jfm.2024.977}}.

\bibitem{ohana2024well}
R.~Ohana, M.~McCabe, L.~Meyer, R.~Morel, F.~Agocs, M.~Beneitez, M.~Berger, B.~Burkhart, S.~Dalziel, D.~Fielding, D.~Fortunato, J.~Goldberg, K.~Hirashima, Y.-F. Jiang, R.~Kerswell, S.~Maddu, J.~Miller, P.~Mukhopadhyay, S.~Nixon, J.~Shen, R.~Watteaux, B.~R{\'e}galdo-Saint~Blancard, F.~Rozet, L.~Parker, M.~Cranmer, S.~Ho, {The Well}: a large-scale collection of diverse physics simulations for machine learning, in: Advances in Neural Information Processing Systems, Vol.~37, 2024, pp. 44989--45037.
\newblock \href {https://doi.org/10.52202/079017-1430} {\path{doi:10.52202/079017-1430}}.

\bibitem{welch1967spectrum}
P.~D. Welch, The use of fast {Fourier} transform for the estimation of power spectra: A method based on time averaging over short, modified periodograms, IEEE Transactions on Audio and Electroacoustics 15~(2) (1967) 70--73.
\newblock \href {https://doi.org/10.1109/TAU.1967.1161901} {\path{doi:10.1109/TAU.1967.1161901}}.

\bibitem{kornblith2019similarity}
S.~Kornblith, M.~Norouzi, H.~Lee, G.~Hinton, \href{https://proceedings.mlr.press/v97/kornblith19a.html}{Similarity of neural network representations revisited}, in: Proceedings of the 36th International Conference on Machine Learning, Vol.~97 of Proceedings of Machine Learning Research, PMLR, 2019, pp. 3519--3529.
\newline\urlprefix\url{https://proceedings.mlr.press/v97/kornblith19a.html}

\end{thebibliography}
\end{document}